\documentclass[smallcondensed]{svjour3}     % onecolumn (ditto)
\smartqed  % flush right qed marks, e.g. at end of proof
\usepackage{graphicx}
\usepackage{amsmath}
\usepackage[retainorgcmds]{IEEEtrantools}
\usepackage{mathptmx}      % use Times fonts if available on your TeX system
\journalname{Journal of Mathematical Biology}
\begin{document}
\title{Virus Antibody Dynamics in Primary and Secondary Dengue Infections}
\titlerunning{Dengue Virus and Antibody Dynamics}       
\author{Tanvi P. Gujarati \and G. Ambika }
\institute{Tanvi P. Gujarati \at Indian Institute of Science Education and Research,Tvm, 
 Thiruvananthapuram-695016, Kerala, India, \\\email{tanvi@iisertvm.ac.in}            
              \and
           G. Ambika \at Indian Institute of Science Education and Research,Pune, Pune-411021, Maharashtra, India, \\\email{g.ambika@iiserpune.ac.in}             
}

\date{Received: date / Accepted: date}
\maketitle
\begin{abstract}
Dengue viral infections show unique infection patterns arising from its four serot-
ypes, (DENV-1,2,3,4). Its effects range from simple fever in primary infections to potentially fatal secondary infections. We analytically and numerically analyse virus dynamics and humoral response in a host during primary and secondary dengue infection for long periods using micro-epidemic models. The models presented here incorporate time delays, antibody dependent enhancement (ADE), a dynamic switch and a correlation factor between different DENV serotypes. We find that the viral load goes down to undetectable levels within 7-14 days as is observed for dengue infection, in both cases. For primary infection, the stability analysis of steady states shows interesting dependence on the time delay involved in the production of antibodies from plasma cells. We demonstrate the existence of a critical value for the immune response parameter, beyond which the infection gets completely cured. For secondary infections with a different serotype, the homologous antibody production is enhanced due to the influence of heterologous antibodies. The antibody production is also controlled by the correlation factor, which is a measure of similarities between the different DENV serotypes involved. Our results agree with clinically observed humoral responses for primary and secondary infections.
%\PACS{05.45.-a}
\end{abstract}
\section{Introduction}
\label{Intro}
Dengue is one of the most serious and most difficult to understand diseases infecting humans. According to WHO estimates, 50 to 100 million infections occur each year leading to 500,000 hospitalizations and 20,000 deaths \cite{WHO09}. Until recently dengue was considered to be a disease of the tropics, but now it has spread its domain of infection to temperate regions as well, primarily boosted by global warming \cite{SciD98}. Dengue is transmitted to humans through the bite of infected \textit{Aedes aegypti} and \textit{A. albopictus} mosquitoes. It is understood that four closely related serotypes of DENV exist viz. DENV-1, DENV-2, DENV-3 and DENV-4 \cite{Lin01,Hal88} and these four serotypes cause infections of varying severity in humans. The infected individual usually suffers from acute febrile illness called Dengue Fever (DF) which is cleared by a complex immune response in a short time of approximately 7 days after onset of fever. There are more severe manifestations of this disease like dengue haemorrhagic fever (DHF) and dengue shock syndrome (DSS). Often without immediate and necessary proper treatment DHF/DSS can be fatal \cite{Gib02,Hal07,Rig98}. WHO has recently proposed a new classification for dengue based on disease severity \cite{WHOclass}. We note that, though there is a huge effort going on to develop an effective vaccine against dengue infections, commercial dengue vaccines are not yet available \cite{Mur11,Kin01}. In this context it is important to understand the biological mechanisms and dynamical processes involved during this infection. Also these complex non-linear biological processes lead to dynamic models that are interesting for their varied and rich dynamics.

The epidemiology of dengue in different populations have been studied previously using improved or extended versions of the basic SIR model \cite{Die75,Est98,Est99,Est03,Fen97,Der06,Gar08,Nur07}. Here we present a mathematical model on a cellular scale that incorporates time delays arising from multistage complex processes in immune response. Similar models describing the cell-virus interaction dynamics have been studied in other contexts like HIV, hepatitis or influenza A \cite{Now00,Amb09,Bea11}. However, we note that micro-epidemiological studies concerning of DENV are very few, one such study reported recently involves the T-cell immune response \cite{Nur09}. 

To develop a model for Dengue infection at the micro-epidemic level, the multistage cellular processes that occur during the infection are to be considered in detail. It is understood that once virions enter into the body, they infect macrophages, monocytes, dendritic cells, mast cells and hepatocytes \cite{Jin04,Kli89,Tas03,Wu00}. The virions start multiplying in the infected cells and  burst out of them in large numbers, thus aggravating the infection.  The immune response of the body tries to curb this infection. Generally, adaptive immune response is composed of the cell-mediated immune response and humoral immune response, both of which are responsible to clear infection and provide immunity \cite{Jan01}. This is true with dengue pathogenesis as well \cite{Mur11,Ohm97}. In the case of dengue though, among the different complex immune response mechanisms, the humoral immune response has been shown to play a more prominent role \cite{Hal88,Hal07,Mur11,Hal09,Jen08,Cha78,Run95}. In this mechanism, the B-cells that come into contact with the antigens present on the virus particles start the process of producing antibodies. The antibodies thus produced to complement the viral antigens neutralise the virus particles, making them non-infectious. These virus-antibody aggregates are then chemically degraded and destroyed in the macrophages (note that macrophages are also DENV targets). The external manifestations of this whole process is the acute febrile illness that gets cured within 7-14 days even in the absence of medication \cite{Mur11,Sco10}. If a person is exposed to DENV of any serotype for the first time, it is termed as primary infection. It is observed that infection with a serotype provides long-term immunity against that serotype, i.e. recurrence of dengue infection with the same serotype is not observed \cite{Mur11}. This is because the \textquoteleft memory\textquoteright of the previous infection suppresses the spread of virions without much delay \cite{Hal07,Jan01,Mar97}. 

In this work we propose a within-host dynamical model that involves humoral or antibody mediated immune response to describe primary dengue infection. We also introduce intra-cellular time delays that come from the various steps involved during antibody production \cite{Cha78,Run95}. We carry out a detailed stability analysis followed by numerical analysis to identify the relevant range of parameters that correspond to several possible outcomes of the infection. Our results indicate that there exists a critical value for the immune response parameter, above which the disease free state is always stable. However, large time delay in antibody production can upset this and lead to recurring spikes of virus counts. 

Further, if an individual who has already undergone primary infection is again exposed to DENV of a different serotype, it is called secondary infection. It is generally observed that complicated and fatal conditions of DSS/DHF occur in patients suffering from secondary infection. The new DENV serotype is structurally similar to the old one, hence it leads to production of antibodies that complement the old serotype (i.e. cross immunity) along with new antibodies that are made to combat the new serotype. Thus, there are mainly two relevant antibodies now, one from the old infection which are heterologous to the new serotype and the new homologous antibodies from the present infection. Both these antibodies have the capacity to bind to virions and neutralise them. Once neutralised the antibody-virus complex is taken to the macrophages. This time though, on entering the macrophage, the heterologous antibodies are released from the virus due to their low affinity for this new serotype. This makes a fraction of the virus particles bare and  ready for infecting the macrophage. Thus, the antibodies produced against the new serotype protects as well as carries the virus to their targets depending on the affinity between the antigen and antibody. This process is called the Antibody Dependent Enhancement (ADE) of infection \cite{Mur11,Por86}. ADE has been the most dominant hypothesis for decades that could explain the enhanced severity in secondary infections \cite{Mur11,Ohm97,Hal70}. It must be mentioned that there are other proposed theories like the theory of original antigenic sin for the same \cite{Rot11,Jul11,Noi08}, but we focus on the modelling of ADE mechanism in the present work. Clinically it is observed that during the secondary infection the viral counts are higher than that of the primary infection but the virus gets cleared from the body in about the same time\cite{Hal07,Mur11,Hal70,Vau00}. The other important feature of secondary infections is a very high concentration of antibodies in the bodies of patients. It is postulated that this high load of antibodies could further lead to severe conditions described by DSS/DHF \cite{Hal07,Mur11,Ohm97}. We thus see that it is the immune response which is responsible for controlling the infection and providing immunity as well as for stimulating reactions in the body leading to severe symptoms as seen in DHF/DSS \cite{Hal07,Mur11}. Thus a model for secondary infection should capture the observed features of ADE of infection and the enhanced antibody production.

We extend the primary infection model to a 6-dimensional secondary infection model with heterologous and homologous antibodies based on the above described mechanism. To the best of our knowledge, it is the first attempt at modelling the within-host processes during a secondary dengue infection. In addition to time delays, our model introduces a correlation factor between the viral serotypes to bring in the effect of relatedness of the different DENV serotypes (due to their antigenic similarities) into the dynamics, which is very important in the context of secondary infection. Also the dynamical switch introduced in the model takes care of the influence of homologous antibody concentrations present in the body from the primary infection. With these additional parameters, our model accounts for the Antibody Dependent Enhancement of virus as well as the increased antibody counts observed clinically during secondary dengue infection.

We present the model for primary infection in section \ref{sec:1} and its analytical treatment in section \ref{sec:2}. The model for the secondary infection is introduced in section \ref{sec:4} followed by its numerical analysis in section \ref{sec:5}. The discussion on our model and concluding remarks are given in the last section. 

\section{Model describing primary dengue infection}
\label{sec:1}
\begin{figure}
\begin{tabular}{cc}
\includegraphics[width=0.45\textwidth]{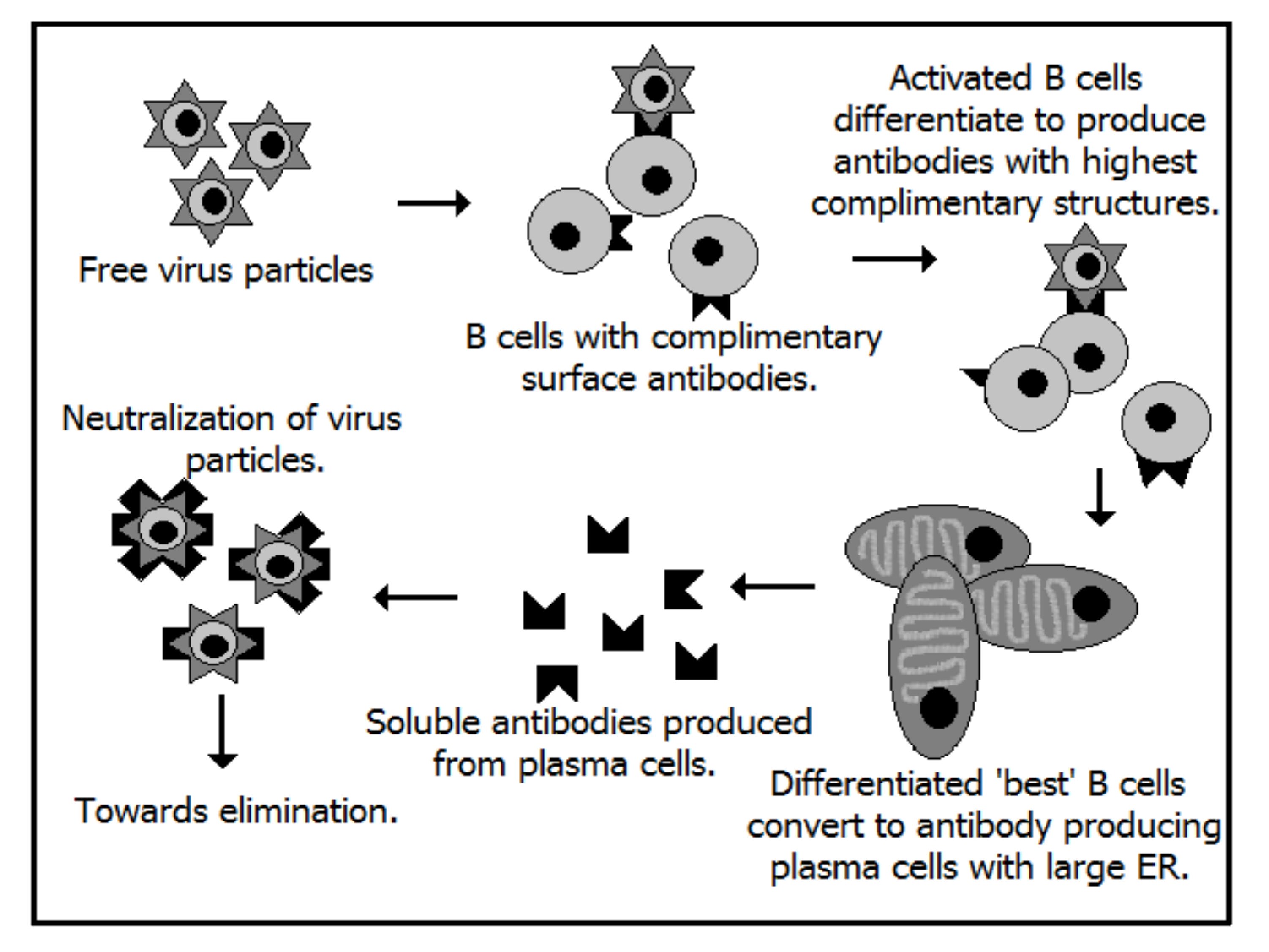} &
\includegraphics[width=0.45\textwidth]{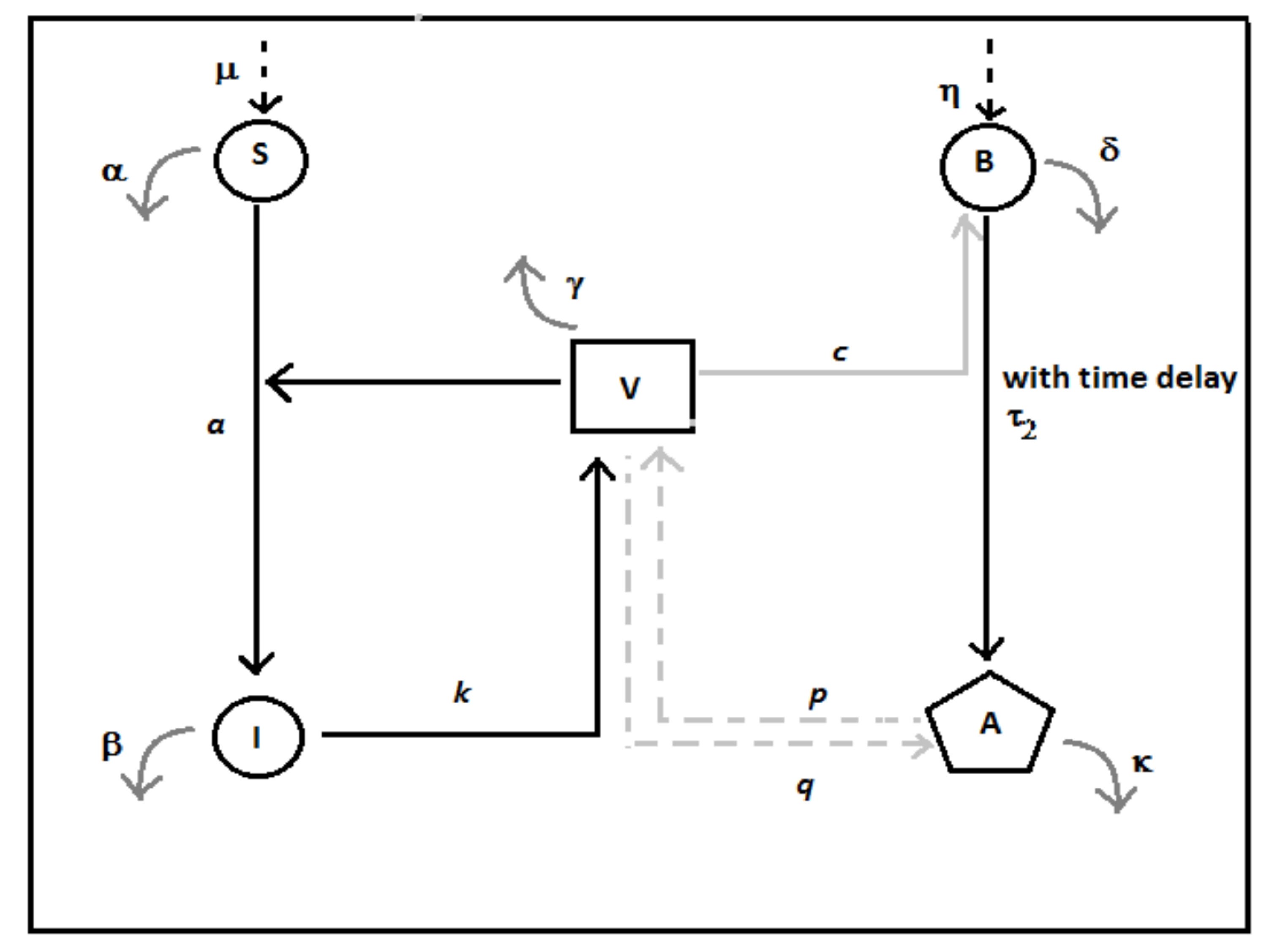} \\
\mbox{\bf 1a} & \mbox{\bf 1b}
\end{tabular}
\caption{ \label{fig:fig1} Humoral immune response in primary dengue infection. Fig.1a Cartoon representation of virus-antibody interactions. Fig.1b Schematic of the mathematical model. Dashed black arrows stand for intrinsic growth rate. Solid black arrows depict conversion and infection. Solid straight grey arrow shows activation rate. Curved grey arrows show intrinsic death rates and dashed grey arrows show elimination rates.}
\end{figure}

In this section, we present the mathematical model describing the primary infection as a 5-D model that takes into consideration the viral dynamics and humoral immune response as described in the introduction. It is built on the basic 3-D model with healthy cells, infected cells and virus particles proposed by May and Nowak \cite{Now00} with humoral response involving B cells and antibodies added to it. The pictorial representation of the B-cell dependent humoral response is given in Fig.\ref{fig:fig1}a and a schematic representation of the model described below is shown in Fig.\ref{fig:fig1}b

The equations describing the model are given by:
\begin{eqnarray}  
  \dot{S} & = & \mu - \alpha S - aSV \nonumber \\ 
  \dot{I} & = & aSV - \beta I \nonumber \\
  \dot{V} & = & kI - \gamma V - pAV \nonumber \\
  \dot{B} & = & \eta - \delta B + cBV \nonumber \\
  \dot{A} & = & fH(t-\tau_{1})B(t-\tau_{2}) - qAV - \kappa A
\label{eq:primarymodel}      
\end{eqnarray}
where the relevant variables are: S-healthy cells (e.g. among monocytes, macrophages, dendritic cells, mast cells or hepatocytes),  I-infected cells, V-Dengue virus particles, B- B lymphocytes and A-neutralizing antibodies.
The description of the parameters controlling the dynamics are given in Table 1.
\begin{table}
\caption{Description of the parameters used in the model.}
\label{tab:1}       
\begin{tabular}{ll}
\hline\noalign{\smallskip}
 Parameter Symbols & Parameter Description \\
\noalign{\smallskip}\hline\noalign{\smallskip}
 $\mu$ & Production rate of healthy cells \\ 
 $\alpha$ & Death rate of healthy cells \\ 
 $a$ & Rate at which healthy cells are converted to infected cells  due to their interaction\\
 & with virus particles \\ 
 $\beta$ & Death rate of infected cells \\ 
 $k$ & Burst rate of virus particles \\ 
 $\gamma$ & Rate at which virus particles degrade \\ 
 $p$ & Rate at which virus particles are neutralised by antibodies \\ 
 $\eta$ & Production rate of B-lymphocytes \\ 
 $\delta$ & Death rate of B cells \\ 
 $c$ & Rate at which B cells are stimulated  by virus particles \\ 
 $f$ & Rate at which stimulated B cells (Plasma cells) produce antibodies \\ 
 $q$ & Rate at which antibody-virus complex affects the antibody growth. \\ 
 $\kappa$ & Rate at which free antibodies degrade. \\ 
\noalign{\smallskip}\hline
\end{tabular}
\end{table}

The concept of Heaviside step function, $ H $, and delay in antibody production is similar to the ones used earlier in immune response models \cite{Fow81,Dib77}. Our model includes two time delays $\tau_{1}$ and $\tau_{2}$. For an infection beginning at time t=0, the total time delay $\tau_{1}$, introduced through the Heaviside step function, is the time period that is required for the first production of antibodies after the virus and B-lymphocytes interact. This process involves two significant biological steps: conversion of B-cells to plasma cells and production of antibodies from plasma cells. On coming in contact with the virions for the first time, the B-cells have to undergo multiple differentiations before they can be transformed into the plasma cells or B-cells capable of producing antibodies specific to the attacking viral antigen \cite{Jan01}. This delay is biologically significant since production of plasma cells (denoted by the same variable as B-cells in our model \cite{Kle80}) after the virions have interacted with the B-lymphocytes is a complex process involving multiple steps. The second part of the total delay, more specifically, time delay $\tau_{2}$ introduced in our model, takes into consideration the time required to produce antibodies from the plasma cells. This would mean that at any time $t$, the model will consider the plasma cell concentration present at time $(t - \tau_{2})$ for the production of antibodies. It must be noted that the time delay $ \tau_{1} $ includes the time delay $ \tau_{2} $. 

For this model the reproduction ratio, $R_0$, calculated by using the next generation method \cite{Die00} is:
\begin{equation}
R_{0} = \frac{a \mu k}{\alpha \beta \gamma}
\end{equation}

The detailed analysis of the possible steady states based on the model described by equations (\ref{eq:primarymodel}) is given in the next section.

\section{Analysis of equilibrium states and their stability}
\label{sec:2}
Primary dengue infection is known to clear from the host completely with the virus getting removed totally from the body after infection and the antibodies to the primary infection persist in the body for a long time \cite{Mur11}. Hence we are interested in looking for infection free solutions and their stability over a long time. For this we solve the modelling equations for all the equilibrium values. Then a detailed analysis of their stability is carried out to identify the relevant parameter values that lead to infection free steady states or infectious states after a long time. 

The equilibrium values denoted with  $'*'$ are obtained from equation (\ref{eq:primarymodel}) as :
\begin{subequations}
\label{eq:equilibrium}
\begin{eqnarray}
  S^{*} & = & \frac{\mu}{(\alpha + aV^{*})}  \\  
  I^{*} & = & \frac{aS^{*} V^{*}}{\beta}  \\
  B^{*} & = & \frac{\eta}{(\delta - cV^{*})}\\
  A^{*} & = & \frac{f B^{*}}{( qV^{*} + \kappa)}  
\end{eqnarray}
\end{subequations}
and  \\
\begin{equation}
V^{*}(p_{3}V^{*3} + p_{2}V^{*2} + p_{1}V^{*} + p_{0}) = 0
\label{eq:cubic}
\end{equation}
where\\
\begin{subequations}
\label{eq:coefficients}
\begin{eqnarray}
p_{3} & = & qca \gamma \\
p_{2} & = & [a \gamma (c \kappa - q\delta)-qc \alpha \gamma(R_{0} - 1)] \\
p_{1} & = & [\alpha \gamma(q \delta - \kappa c)(R_{0}-1)- \kappa \gamma \delta a -apf \eta] \\
p_{0} & = & \alpha [\kappa \delta \gamma(R_{0} - 1)-fp \eta] 
\end{eqnarray}
\end{subequations}

We find equation (\ref{eq:cubic}) has one solution  $V^{*} = 0$  another $V^{*} \neq 0$  which can be obtained from the equation

\begin{equation}
p_{3}V^{*3} + p_{2}V^{*2} + p_{1}V^{*} + p_{0} = 0
\label{eq:Vnonzero}
\end{equation}

The existence of a positive solution for this equation will depend on the values of the parameters.  

To study the stability of the equilibrium state, we linearise the system about the equilibrium points and the corresponding Jacobian matrix given as the following \cite{Lak10} 
\begin{equation}
 J = \left[ \begin{array}{ccccc}
-(\alpha+aV^{*}) & 0 & -aS^{*} & 0 & 0\\
aV^{*} & -\beta & aS^{*} & 0 & 0\\
0 & k & -(\gamma+pA^{*}) & 0 & -pV^{*}\\                                             
0 & 0 & cB^{*} & -(\delta-cV^{*}) & 0\\
0 & 0 & -qA^{*} & fe^{-\lambda\tau_{2}} & -(\kappa+qV^{*})\\
\end{array} \right] 
\label{eq:jacobian}
\end{equation} 
Here $\lambda$ stands for the eigenvalues of the Jacobian matrix $J$ given above.
If the real parts of all the eigenvalues obtained by solving the characteristic equation of $J$ are negative, it implies that the system is stable in that equilibrium state. 

On substituting the equilibrium values of $S^{*}$, $B^{*}$ and $A^{*}$ in terms of $V^{*}$ from equation (\ref{eq:equilibrium}), the characteristic equation for the $J$ can be written as
\begin{eqnarray}
\label{eq:chareq}
G(\lambda) & = & \lambda^{5} + G_{4} \lambda^{4} + G_{3} \lambda^{3} + G_{2} \lambda^{2} + G_{1} \lambda + G_{0}\\ \nonumber
&& +\: H_{2}e^{-\lambda \tau_{2}} \lambda^{2} + H_{1}e^{-\lambda \tau_{2}} \lambda + H_{0}e^{-\lambda \tau_{2}} \\ \nonumber
&& =\: 0 \\ \nonumber
\end{eqnarray} 
The details of all the coefficients in terms of the parameters are given in Appendix.

\subsection{Infection free equilibrium state}
\label{sec:3}
We start by discussing the most interesting case i.e. $V^{*}=0$, the infection free equilibrium state. In this case the characteristic equation in equation (\ref{eq:chareq}) can be simplified as (details in Appendix)
\begin{equation}
G(\lambda) = (\alpha + \lambda)(\kappa + \lambda)(\delta + \lambda)(\lambda^{2} + \lambda(\gamma(1 + \frac{fp\eta}{\delta \gamma \kappa}) + \beta) + \beta \gamma(1 + \frac{fp\eta}{\delta \gamma \kappa} - R_{0}))
\label{eq:chareqv}
\end{equation}
On analysing the equation (\ref{eq:chareqv}) we find that it is independent of $\tau_{2}$. The eigenvalues have negative real parts when $R_{0}>1$ and $f>f_{c}$ given in equation (\ref{eq:fcritical}). 
\begin{equation}
\label{eq:fcritical}
f_{c} = (R_{0}-1) \frac{\kappa \delta \gamma}{p \eta} 
\end{equation}

\begin{figure}[h]
\begin{center}
\includegraphics[scale=0.8]{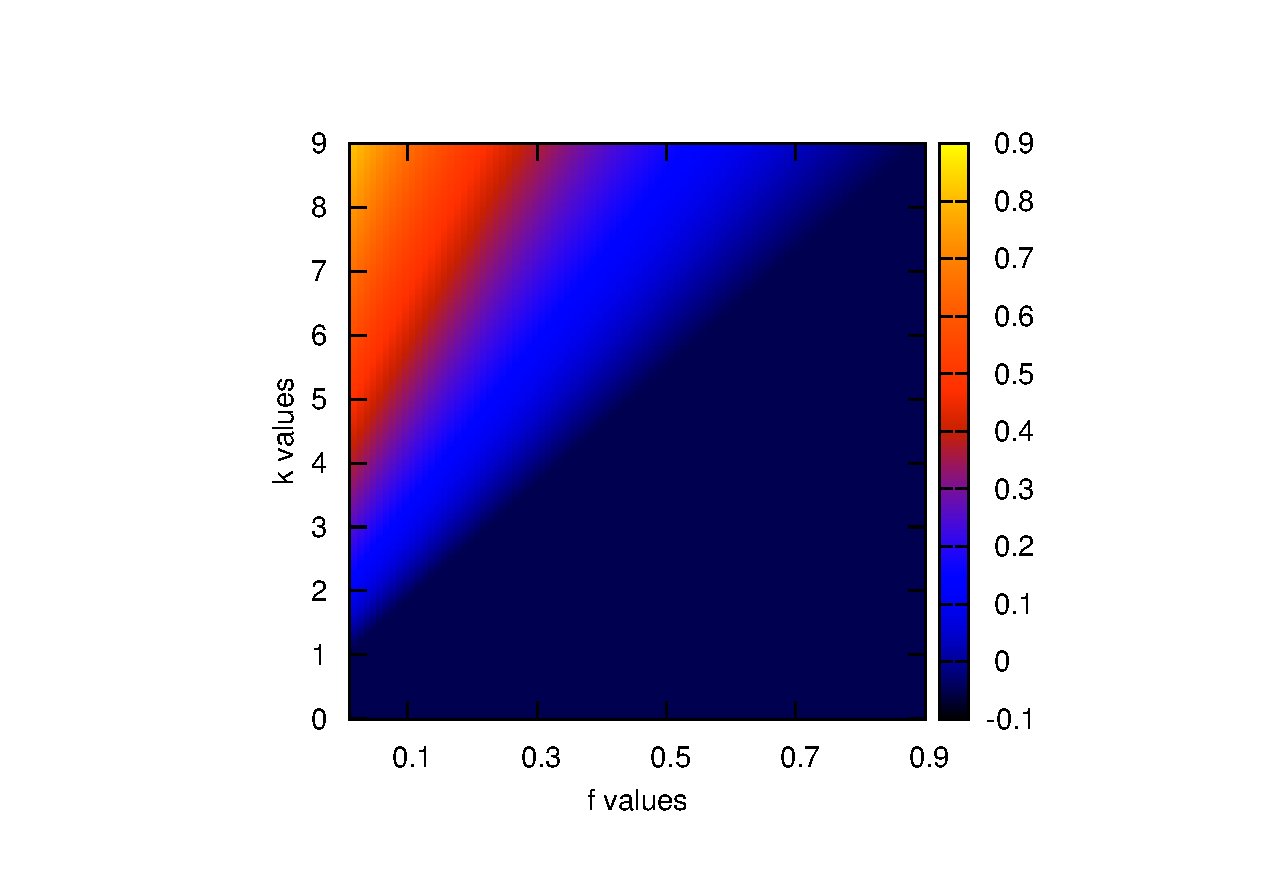}
\end{center}
\caption{\label{fig:figa2} The parameter plane $f$-$k$ indicating stability regions as indexed by the maximum of the real parts of eigenvalues obtained from the characteristic equation (\ref{eq:chareqv}) for $V^{*}=0$. Blackish blue(grey) end of the spectrum denotes negative values and hence implies that the system is in stable equilibrium in those regions. Similarly the yellow(white) end of the spectrum denotes positive values, implying that the system is unstable in those regions. The numerical values of the parameters used are:$ \mu $=10, $ \alpha $=0.05, $ a $=0.001, $ \beta $=0.5, $ \kappa $=0.051, $ \gamma $=0.5, $ p $=0.001, $ \eta $=10, $ \delta $=0.049, $ c $=0.001, $ q $=0.001.}
\end{figure}

We consider the two parameters, $k$, the burst rate of virus ($k$ is directly proportional to the reproduction ratio $R_{0}$) and $f$, the production rate of antibodies as the relevant parameters controlling the dynamics and identify regions of stability  in the parameter plane $f$-$k$. For this the parameter plane is scanned in the steps of 0.01 and colour coded using the maximum value of the real parts of the eigenvalues as index. This is shown in Fig.\ref{fig:figa2}. In this analysis the values of other parameters used are $ \mu $=10, $ \alpha $=0.05, $ a $=0.001, $ \beta $=0.5, $ \kappa $=0.051, $ \gamma $=0.5, $ p $=0.001, $ \eta $=10, $ \delta $=0.049, $ c $=0.001, $ q $=0.001 adapted from models in similar context \cite{Amb09,Bea11}. 

The line separating the blackish blue(dark grey)-blue(whitish-grey) regions in Fig.\ref{fig:figa2} corresponds to the stability boundary of the infection free equilibrium state $V^{*}=0$. Thus for a given $k$, virus burst rate, this line gives the critical value of immune response parameter, $f>f_{c}$ required for eliminating the infection.

We shall now discuss the conditions under which nonzero solution exists for $V^{*}$ by numerically solving equation (\ref{eq:Vnonzero}) and the stability of this equilibrium in the next section.

\subsection{Infectious equilibrium $V^{*} \neq 0$}
\label{sec:12}

\begin{figure}[h]
\begin{center}
\includegraphics[scale=0.8]{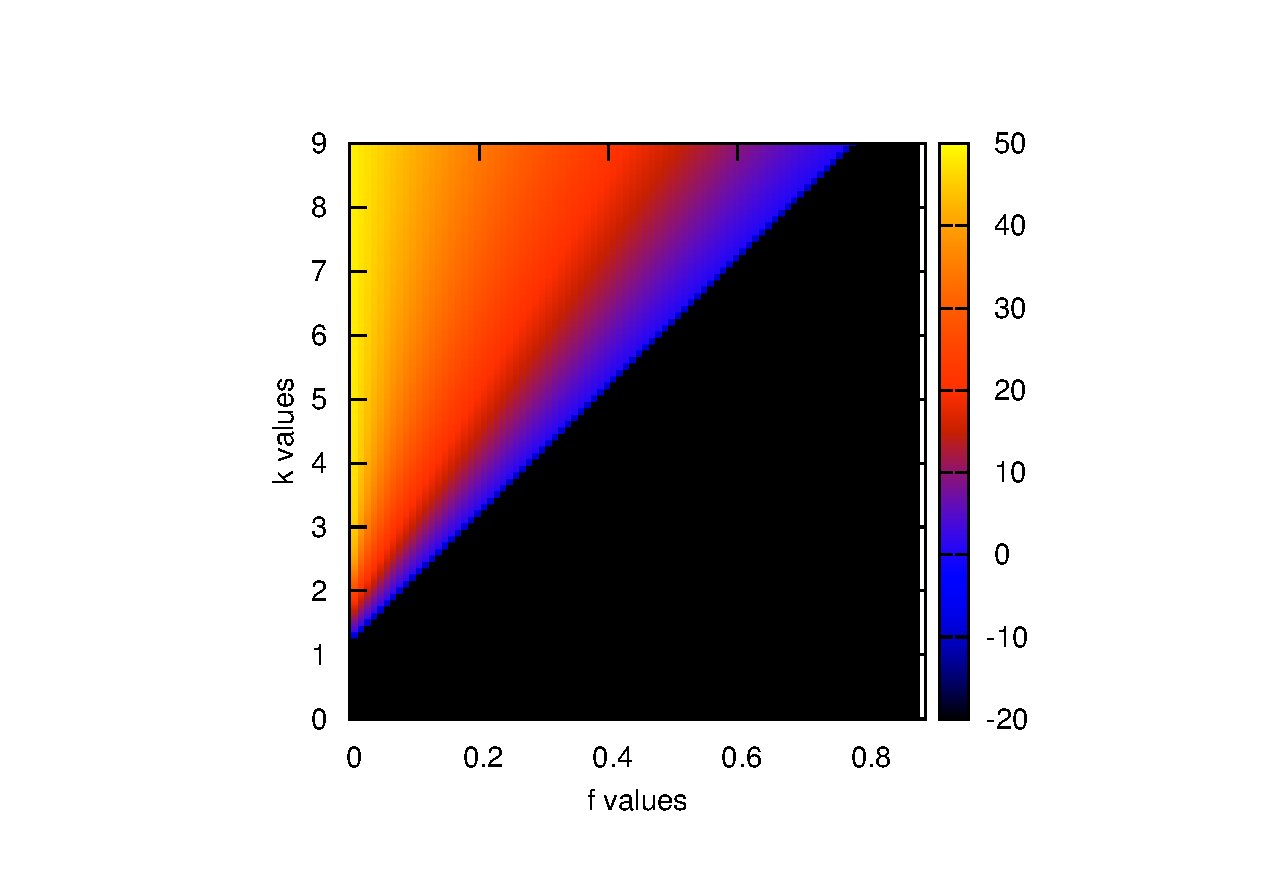}
\end{center}
\caption{\label{fig:figa3} The parameter plane $f$-$k$ indicating the value of $V^{*} \neq 0$ equilibrium solution obtained by numerically solving equation (\ref{eq:Vnonzero}). Blackish (black) end of the spectrum denotes negative values and hence implies that $V^{*} \neq 0$ is biologically irrelevant solution in this region. Similarly, the yellow(white) end of the spectrum denotes positive values, implying that the system has $V^{*} \neq 0$ equilibrium solutions of interest. The numerical values of the parameters used are:$ \mu $=10, $ \alpha $=0.05, $ a $=0.001, $ \beta $=0.5, $ \kappa $=0.051, $ \gamma $=0.5, $ p $=0.001, $ \eta $=10, $ \delta $=0.049, $ c $=0.001, $ q $=0.001.}
\end{figure}

On numerically solving the equation (\ref{eq:Vnonzero}) in the parameter plane ($f-k$) with the same parameter values used in section (\ref{sec:3}), we find that $V^{*} \neq 0$ equilibrium does not exist in the region where $V^{*} = 0$ is stable as is seen in Fig.(\ref{fig:figa3}). The black region corresponds to negative $V^{*}$ values and the positive $V^{*}$ values are colour coded. Note that all the values of $V^{*} \geq \frac{\delta}{c}$ were ignored as they would give unrealistic values for $B^{*}$ as can be infered from equation (\ref{eq:equilibrium})c.

For $V^{*} \neq 0$ the characteristic equation is given by equation (\ref{eq:chareq}). It is clear that the presence of time delay in the dynamics makes the characteristic equation transcendental. Hence the derivation of stability criteria is not straight forward. However we can continue with the analysis by assuming that the eigenvalues is a continuous function of the time delay $\tau_{2}$. We can write $\lambda(\tau_{2}) = \varepsilon(\tau_{2})+i\omega(\tau_{2})$ where $\varepsilon$ and $\omega$ are the real and imaginary parts of $\lambda$ respectively. We analyse the stability in this case in two steps \cite{Lak10}. First we check the stability for $\tau_{2}=0$ and if the state is stable it means that $\varepsilon(\tau_{2})<0$. As $\tau_{2}$ increases, if $\varepsilon(\tau_{2})$ crosses zero at some value of $\tau_{2}$, say $\tau_{2}^{0}$, then it means that the stable state would become unstable at $\tau_{2}^{0}$ and $\lambda(\tau_{2}^{0}) = i\omega_{0}$. If no such value of $\tau_{2}$ exists, then the state will remain stable for all values of $\tau_{2}$ if it was stable for $\tau_{2}=0$. Hence we start our analysis by considering the stable regions for $\tau_{2}=0$ and then continue to find out the conditions where changes in stability occur when $\tau_{2}$ is non-zero.

For $\tau_{2}=0$ , the characteristic equation (\ref{eq:chareq}) becomes:
\begin{equation}
G(\lambda) = \lambda^{5}+G_{4}\lambda^{4} + G_{3}\lambda^{3} + (G_{2} + H_{2})\lambda^{2} + (G_{1} + H_{1})\lambda + (G_{0} + H_{0}) = 0                                  
\label{eq:chareq2}
\end{equation}
\begin{figure}[h]
\begin{center}
\includegraphics[scale=0.8]{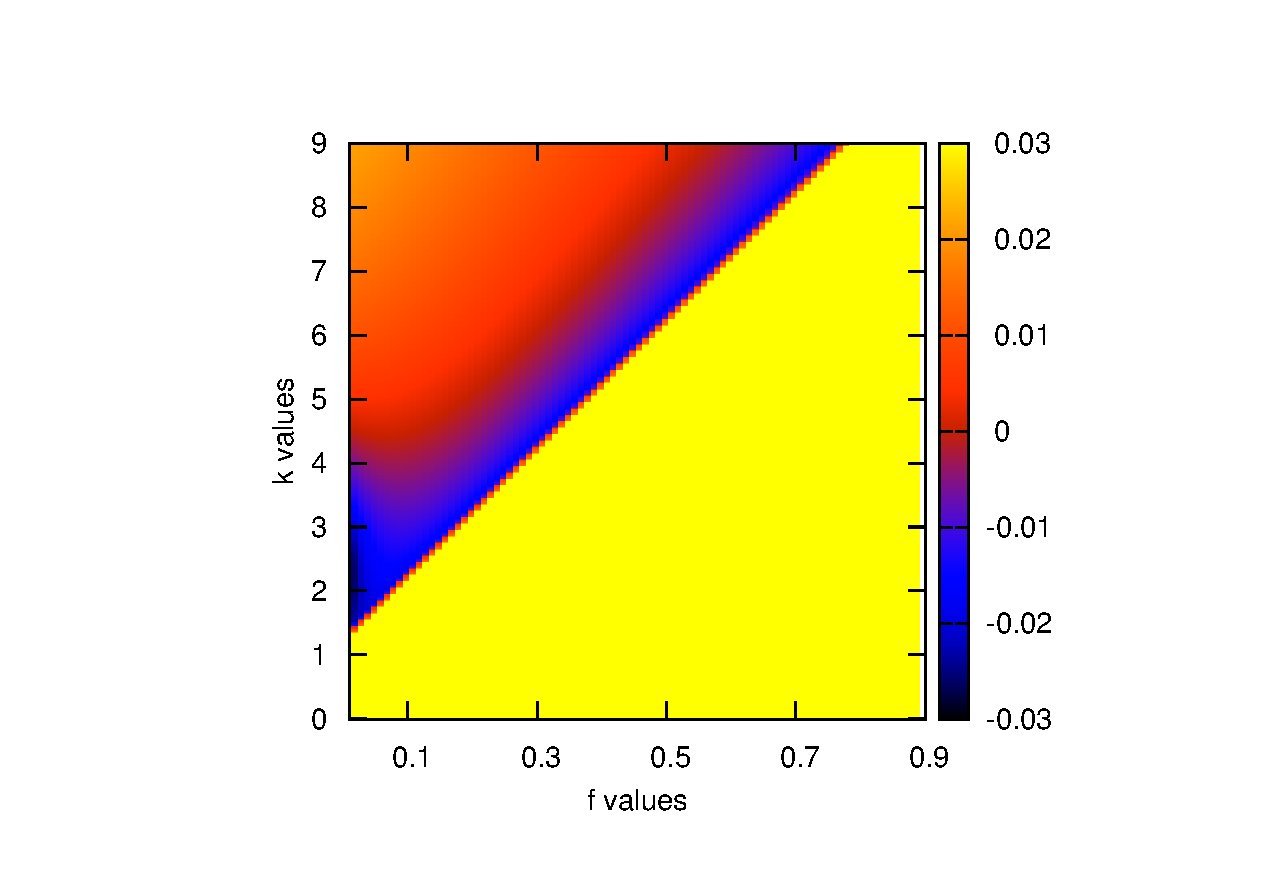} 
\end{center}
\caption{\label{fig:fig2b} The parameter plane  $f$-$k$, indicating stability regions for $V^{*} \neq 0$ as indexed by the maximum of the real parts of eigenvalues obtained from the characteristic equation (\ref{eq:chareq2}) where $\tau_{2}=0$. Blue(black) end of the spectrum denotes negative values and hence implies that the system is in stable equilibrium in those regions. Similarly, the orange(grey) end of the spectrum denotes positive values, implying that the system is unstable in those regions. The region $V^{*} < 0$ and hence is not biologically relevant is given by the yellow (white) region. The numerical values of the parameters used are same as in Fig.\ref{fig:figa2}}
\end{figure}

For $\tau_{2} = 0$, the stability regions for $V^{*} \neq 0$ in the parameter plane ($f-k$) is given                             in Fig.(\ref{fig:fig2b}). The colour coded regions gives the maximum of the real parts of eigenvalues obtained from the equation (\ref{eq:chareq2}) for the corresponding relevant value of $V^{*}$ given in Fig.(\ref{fig:figa3}). We see that regions with blue(black) have negative values and hence are stable regions whereas regions orange(grey) in colour with positive values are unstable. The yellow region corresponds to the region were relevant $V^{*} \neq 0$ does not exist. We also note that the line separating the yellow region from the other regions is similar to the line corresponding to $f_{c}$ observed in Fig.(\ref{fig:figa2}). We can thus conclude that for sufficiently high values of $f$, only $V^{*}=0$ is the relevant and stable equilibrium state. This indicates that if the immune response is sufficiently strong the infection gets cleared and does not recur. 

\begin{figure}[h]
\begin{tabular}{cc}
\includegraphics[scale=0.22]{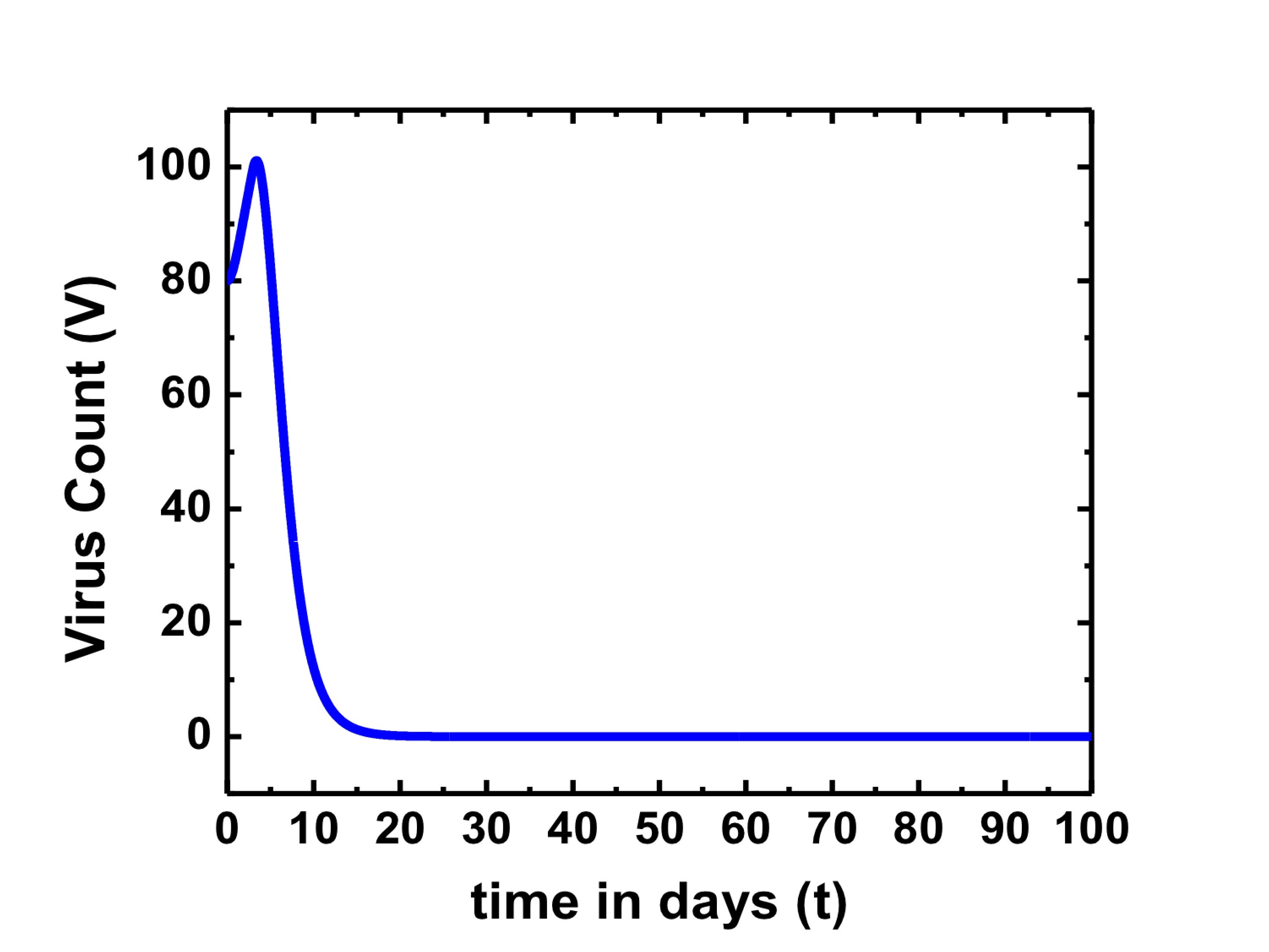}&
\includegraphics[scale=0.22]{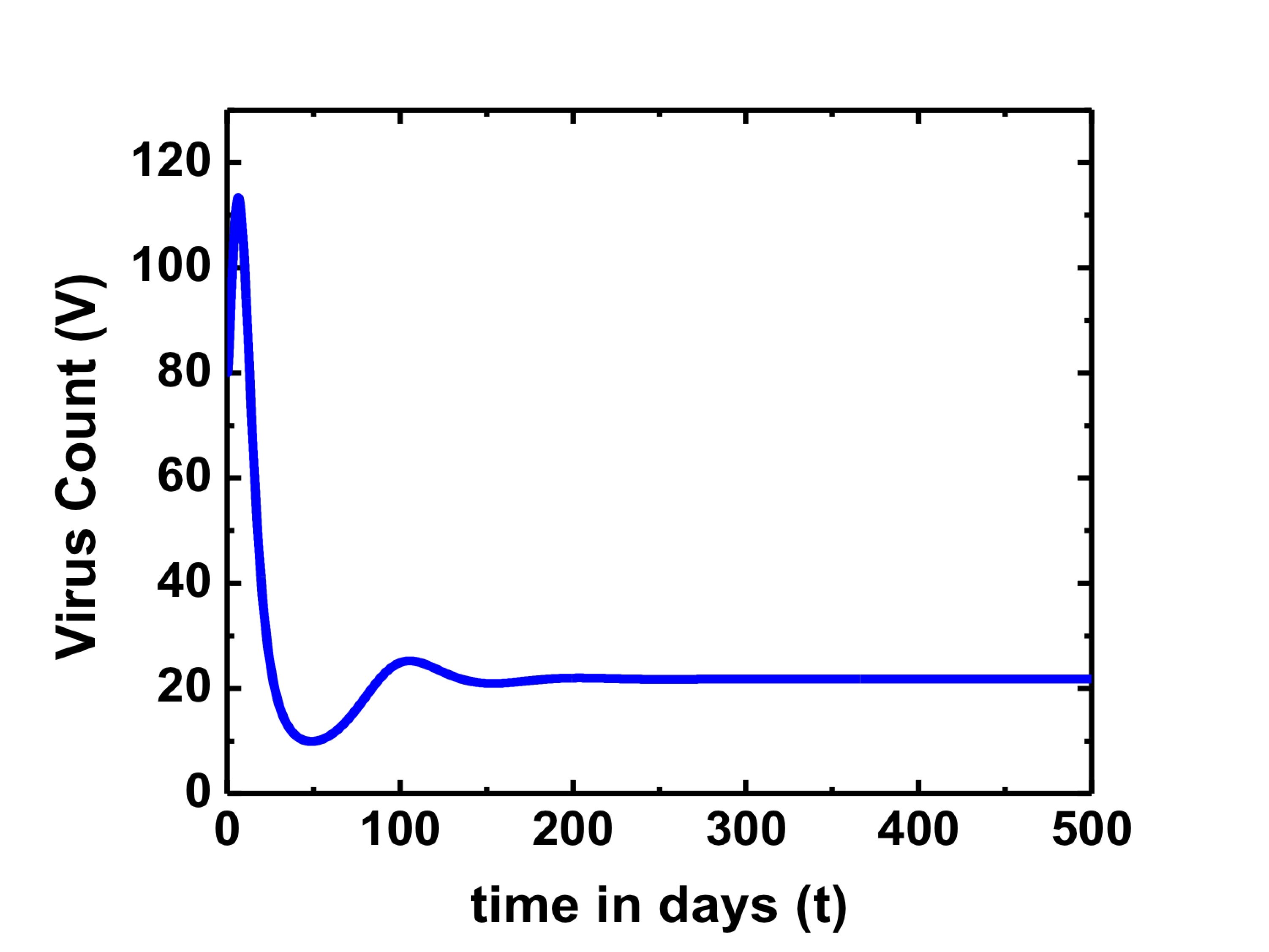} \\ 
\mbox{\bf {5a}} & \mbox{\bf 5b}
\end{tabular}
\begin{center}
\begin{tabular}{c}
\includegraphics[scale=0.22]{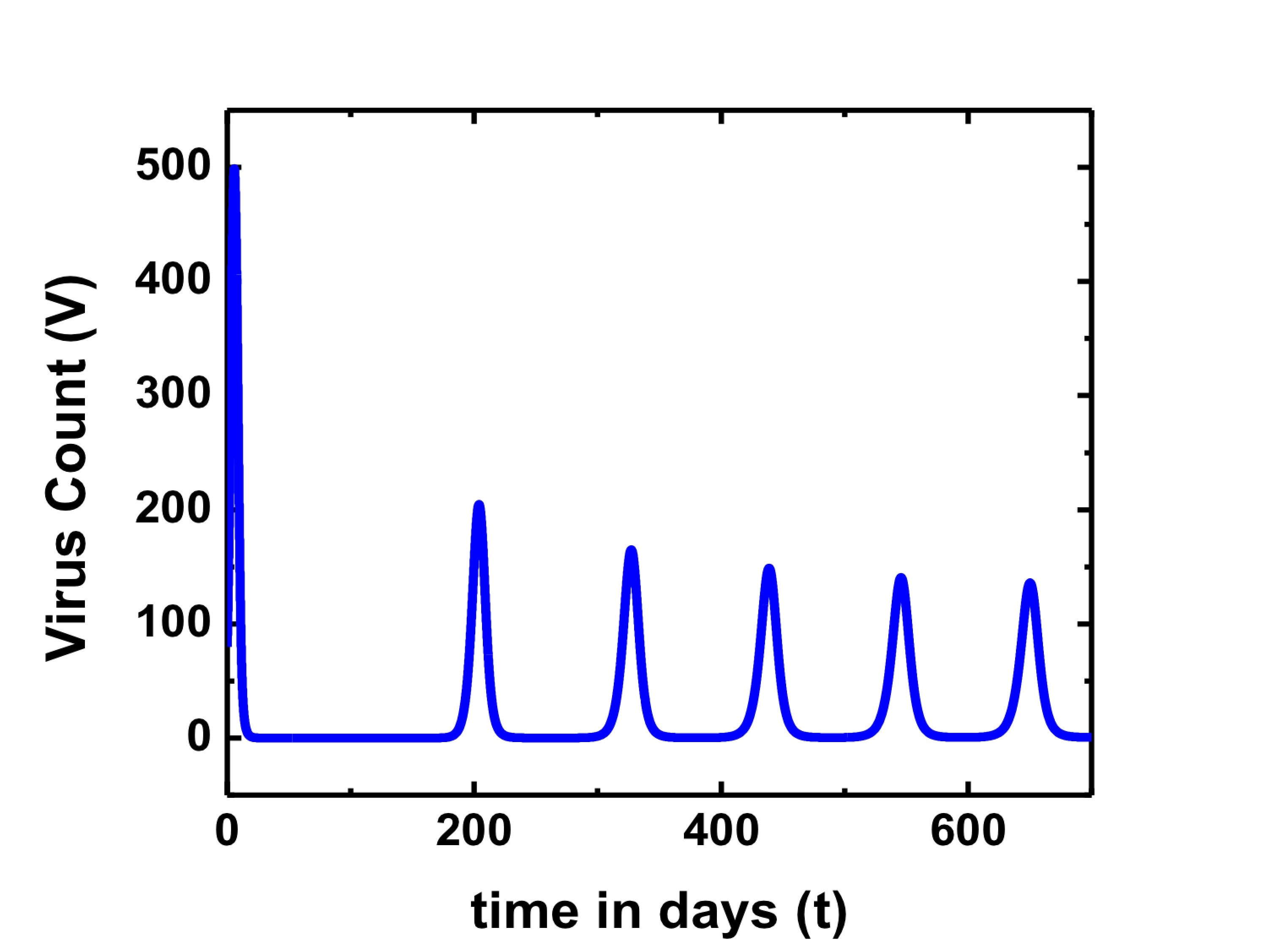}\\
\mbox{\bf 5c}
\end{tabular}
\end{center}
\caption{\label{fig:fig3} Asymptotic dynamics of virus count for specific chosen values of the parameters $f$ and $k$ for $\tau_{2}=0$. 
Fig.4a For $f=0.8$ and $k=2$ corresponds to a point in the stable region with $ V^{*} = 0$. The corresponding time series for the virus particles tend to zero asymptotically. Fig.4b For $f=0.01$ and $k=2$ from the stable region for $V^{*}=V_{1}$ in Fig.\ref{fig:fig2b}, time series for the virus particles settles to non-zero value asymptotically. Fig.4c For $f=0.1$ and $k=5$ from the unstable region, the virus particles show oscillatory behaviour. The parameter values used in the numerical simulations are $ \mu $=10, $ \alpha $=0.05, $ a $=0.001, $ \beta $=0.5, $ \kappa $=0.051, $ \gamma $=0.5, $ p $=0.001, $ \eta $=10, $ \delta $=0.049, $ c $=0.001, $ q $=0.001, $ \tau_{1} $=3 days along with the initial conditions $(S,I,V,B,A)$ = (200,20,80,200,0).}
\end{figure} 

A direct numerical simulation of the equations of the model for illustrative sets of parameter values support the conclusions from the stability analysis given above. The time series of virus particles corresponding to typical regions in the parameter plane showing different asymptotic dynamics are obtained by numerically solving the modelling equations (\ref{eq:primarymodel}) for $\tau_{2}=0$ using the Runge-Kutta method. Here we use initial conditions $(S,I,V,B,A)$ = (200,20,80,200,0) and the parameter values given as:$ \mu $=10, $ \alpha $=0.05, $ a $=0.001, $ \beta $=0.5, $ \kappa $=0.051, $ \gamma $=0.5, $ p $=0.001, $ \eta $=10, $ \delta $=0.049, $ c $=0.001, $ q $=0.001, $ \tau_{1} $=3 days adapted from reported studies in similar context \cite{Amb09,Bea11,Agu94}. We find that for $f=0.8$ and $k=2$, the virus count goes down to zero within 14 days and remains so asymptotically as given in Fig.\ref{fig:fig3}a. For $f=0.01$ and $k=2$, a point from the stable region of $V^{*} \neq 0$ (Fig.\ref{fig:fig2b}), the numerical analysis gives that the virus count goes to a steady non-zero value. It is seen in Fig.\ref{fig:fig3}b. For $f=0.1$ and $k=5$ from the unstable region of both the steady states, we see oscillatory behaviour for virus count (Fig.\ref{fig:fig3}c).

Thus, we find that for sufficiently high $f$ values, the infection gets cleared and does not recur as is observed in primary dengue infections. Also, the infectious steady state with $V^{*} \neq 0$ can occur when $f$ is small enough, which means the infection will not be cleared. It corresponds to the case of very low immune response with high burst rate for virus. Moreover, oscillatory solutions are possible; they consist spikes of virus counts separated by long intervals of very low count. This would correspond to recurrence of infection after an interval. In this context, we add that in the case of HIV, oscillations and recurrence of infection are clinically reported \cite{Kit11}. In the case of dengue, lack of diagnostic measurements of viremia in patients along with the fact that a natural human system has many other intertwined processes simultaneously functional, not just reaction to dengue, hampers the detection of possible oscillatory solution or infectious steady state which may be present in immunocompromised individuals. Hence detailed clinical measurements are required for establishing their existence. 

\subsection{Effect of Time Delay}

The calculations of section 3.1 establish the stability regions of $V^{*}=0$ for any $\tau_{2}$. However, the stability for $V^{*}\neq0$ states is to be further analysed to take care of the case $\tau_{2}\neq0$.

We proceed to get the conditions under which it will remain stable for all values of $\tau_{2}$, i.e. conditions under which $\lambda = i \omega$ is not possible. On substituting $\lambda = i \omega$ in the characteristic equation (\ref{eq:chareq}) we get;
\begin{flalign}
\label{eq:sincos}
&G(i \omega) = i \omega^{5} + G_{4} \omega^{4}-iG_{3}\omega^{3}-G_{2}\omega^{2} + iG_{1}\omega + G_{0}-H_{2}\omega^{2} (\cos (\omega \tau_{2}) -i\sin(\omega \tau_{2})) & \nonumber \\
&           +\: iH_{1}\omega(\cos(\omega \tau_{2})-i\sin(\omega \tau_{2}))+H_{0}(\cos(\omega \tau_{2}) -i\sin(\omega \tau_{2})) = 0 & \nonumber \\
\end{flalign}

We follow the graphical way of finding dependence of stability on time delay by analysing equation (\ref{eq:chareq}) geometrically for $ \lambda = i \omega $. We write equation (\ref{eq:chareq}) as shown below:

\begin{equation}
 e^{-i\omega\tau_{2}} = \frac{((G_{4}\omega^{4}-G_{2}\omega^{2} + G_{0})+i(\omega^{5}-G_{3}\omega^{3} + G_1\omega))}{((H_{2}\omega ^{2}-H_{0})-iH_{1}\omega)}     
\label{eq:geomchareq}
\end{equation}

\begin{figure}[h]
\begin{center}
\begin{tabular}{c}
\includegraphics[width=0.7\textwidth]{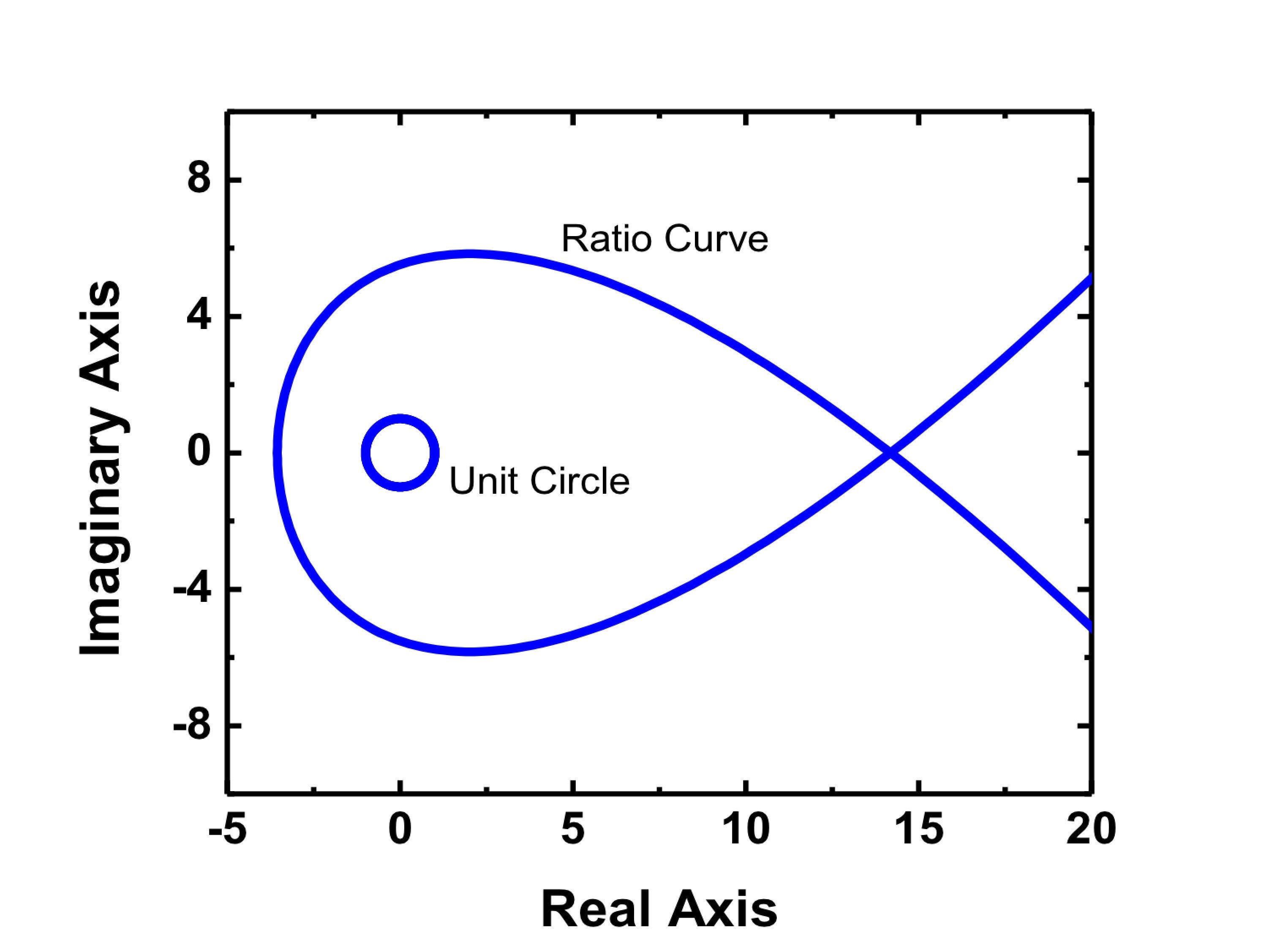} \\
\mbox{\bf 6a} \\
\includegraphics[width=0.7\textwidth]{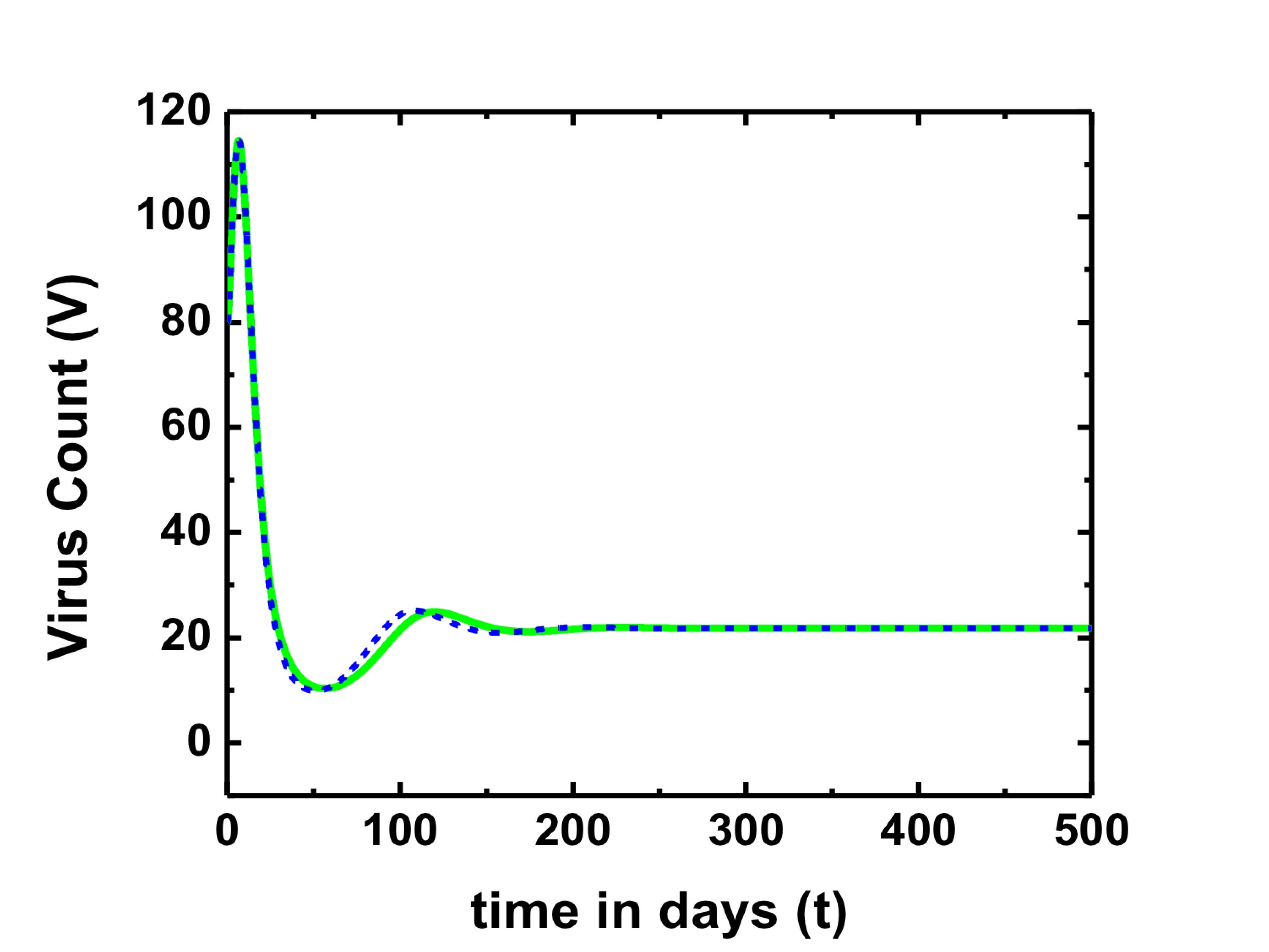} \\
\mbox{\bf 6b} 
\end{tabular}
\end{center}
\caption{\label{fig:fig5} Fig.6a Real and imaginary parts of both sides of equation (\ref{eq:geomchareq}) indicating that stability does not change as unit circle and ratio curve do not intersect, i.e. the equilibrium state $f = 0.01$ and $k = 2$ remains stable for $\tau_{2} \neq 0$. Fig. 6b) Time series for virus particles obtained by numerical simulations for $\tau_{2}$=1 day (dashed curve) and $\tau_{2}$=15 days (solid curve). We see that the virus count stabilises to non-zero counts asymptotically, i.e. the equilibrium point retains its stability for very different non-zero $\tau_{2}$ values. The $\tau_{1}$ value used is 16 days and all other parameter values with initial conditions are given in the text.}
\end{figure}

\begin{figure}[h]
\begin{center}
\begin{tabular}{c}
\includegraphics[width=0.7\textwidth]{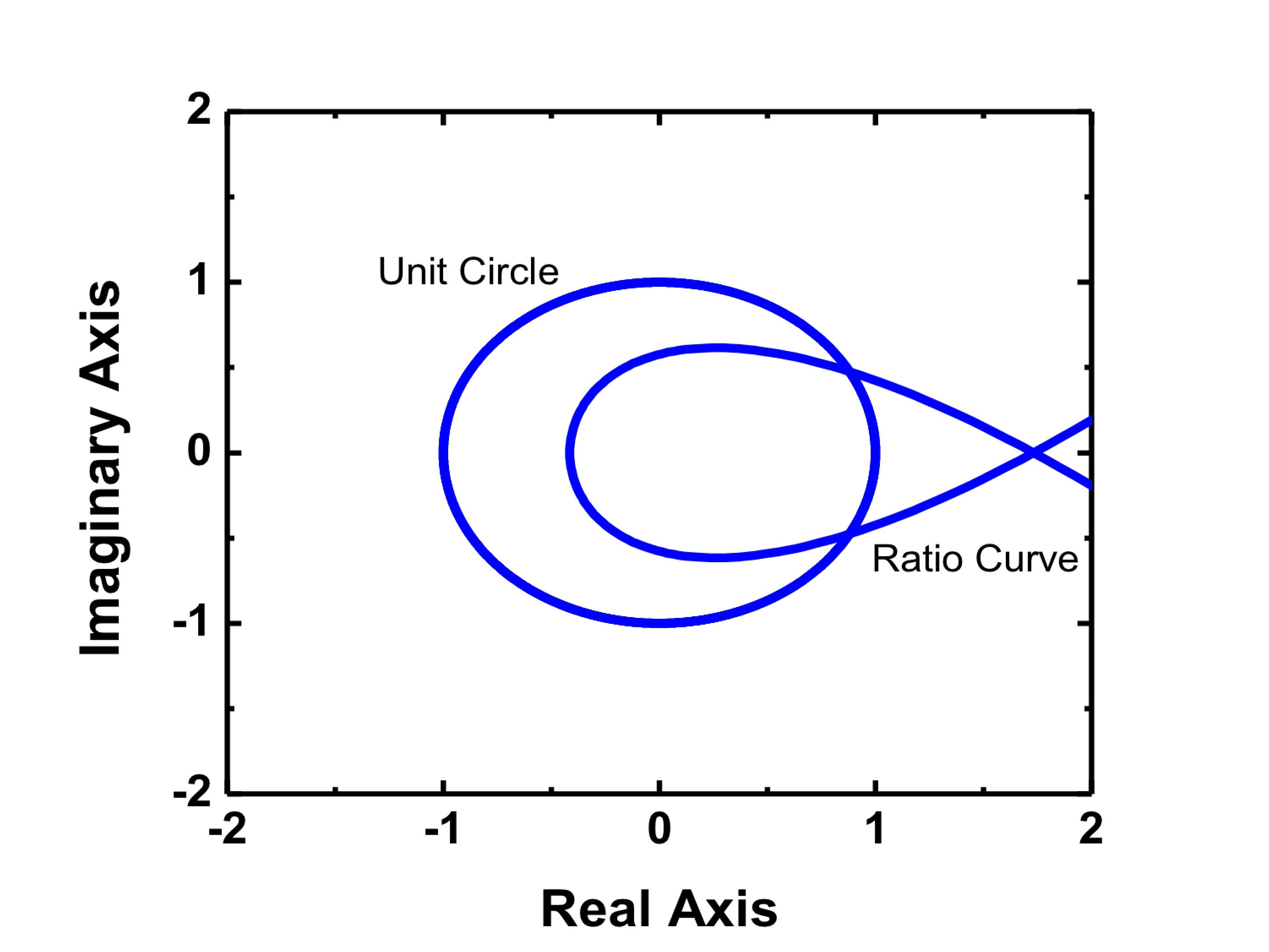} \\ 
\mbox{\bf 7a} \\
\includegraphics[width=0.7\textwidth]{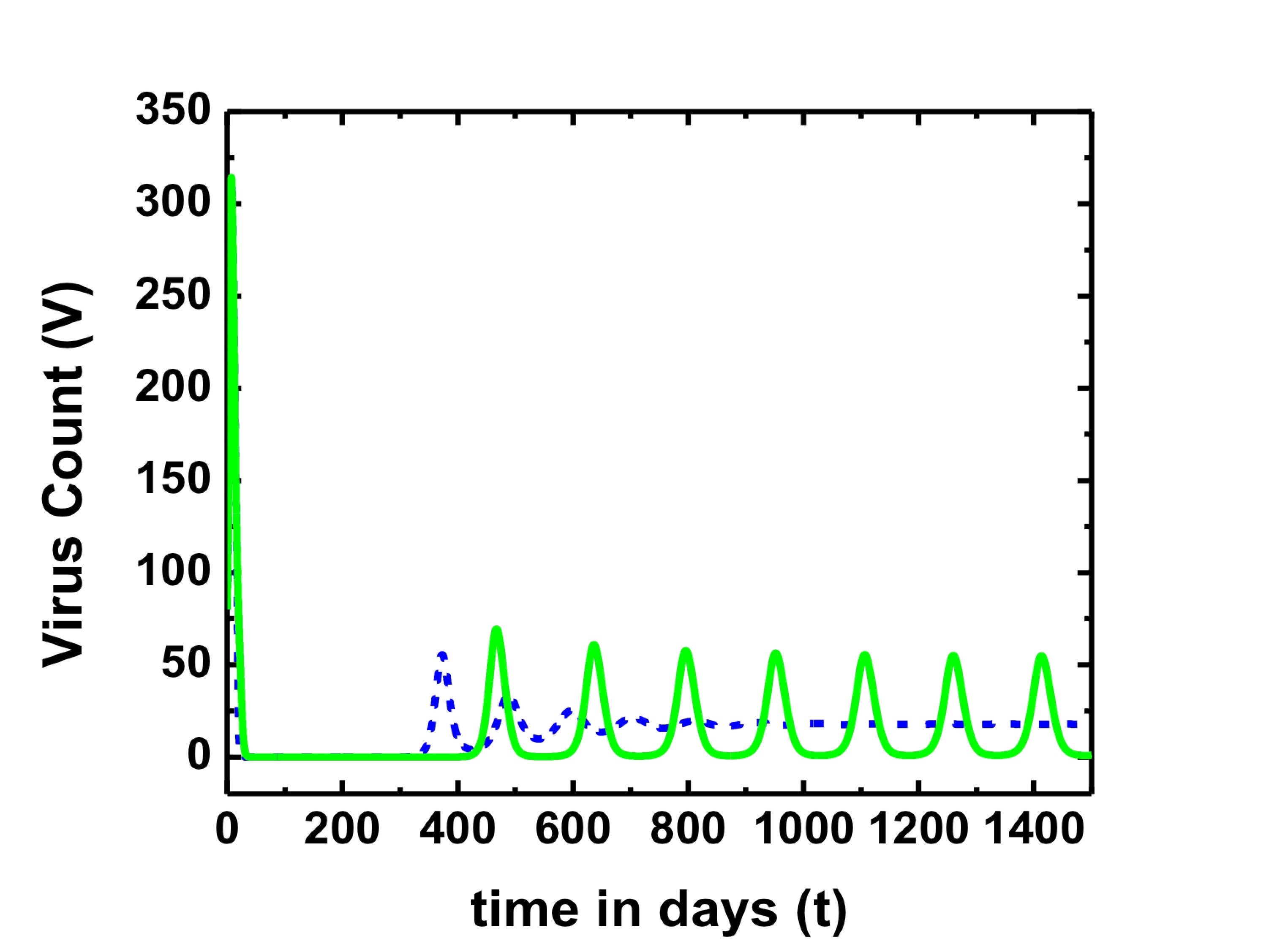} \\
\mbox{\bf 7b} 
\end{tabular}
\end{center}
\caption{\label{fig:fig6} Fig.7a  Real and imaginary parts of both sides of equation (\ref{eq:geomchareq}) indicating that stability changes as unit circle and ratio curve intersect, i.e. the equilibrium state $ f = 0.1$ and $ k = 3.5 $ becomes unstable for $\tau_{2} \neq 0$. Fig.7b Time series for virus particles obtained by numerical simulations for $\tau_{2}$=1 day (dashed curve) and $\tau_{2}$=15days (solid curve). We see that the virus count stabilises asymptotically for smaller values of $\tau_{2} \neq 0$ but becomes oscillatory for higher values of $\tau_{2}$. The $\tau_{1}$ value used is 16 days and all other parameter values with initial conditions are given in the text.}
\end{figure}

The left hand side of the equation (\ref{eq:geomchareq}) defines a unit circle on the Argand plane. The right hand side is called the ratio curve. If the ratio curve intersects the unit circle there is a change of stability. By plotting the real and imaginary parts of both sides of equation (\ref{eq:geomchareq}) simultaneously we can check for the change in stability of the equilibrium point. The point of intersection, if any, is given by equation (\ref{eq:tauvalues}) which is derived from equation (\ref{eq:geomchareq})

\begin{eqnarray}
\tau_{2}^{p} & = & \pm \frac{1}{\omega_{0}} \arccos \frac{[((H_{2}\omega_{0}^{2}-H_{0})(G_{4}\omega_{0}^{4}-G_{2}\omega_{0}^{2}+G_{0}) - H_{1}\omega_{0}(\omega_{0}^{5}-G_{3}\omega_{0}^{3}+G_{1}\omega_{0})]}{[(H_{0}-H_{2}\omega_{0}^{2})^{2}+\omega_{0}^{2}H_{1}^{2})]} \nonumber \\
&& +\: \frac{2p\pi}{\omega_{0}}, \qquad \qquad p = 0,1,2..
\label{eq:tauvalues} 
\end{eqnarray}
We illustrate this analysis by taking the following specific cases:\\
Case1: For $f=0.01$ and $k=2$, which is a point of stable equilibrium for $\tau_{2} = 0$, the plots of the unit circle and ratio curve are shown in Fig.\ref{fig:fig5}a. As the curves do not intersect, the equilibrium point will be stable for all $\tau_{2}$ values. The time series for virus particles obtained by direct numerical analysis with $\tau_{2} \neq 0$ shown in Fig.\ref{fig:fig5}b also supports this. \\
Case2: For $f=0.1$ and $k=3.5$, which was a steady state for $\tau_{2} = 0$, the plots of the unit circle and ratio curve intersect each other as shown in Fig.\ref{fig:fig6}a. Thus, the equilibrium point changes its stability. The time series for the virus particles obtained in Fig.\ref{fig:fig6}b from a direct numerical analysis shows that the solution becomes oscillatory for higher values of $\tau_{2}$. 
\\ 

\begin{figure}[h]
\begin{center}
\includegraphics[width=0.8\textwidth]{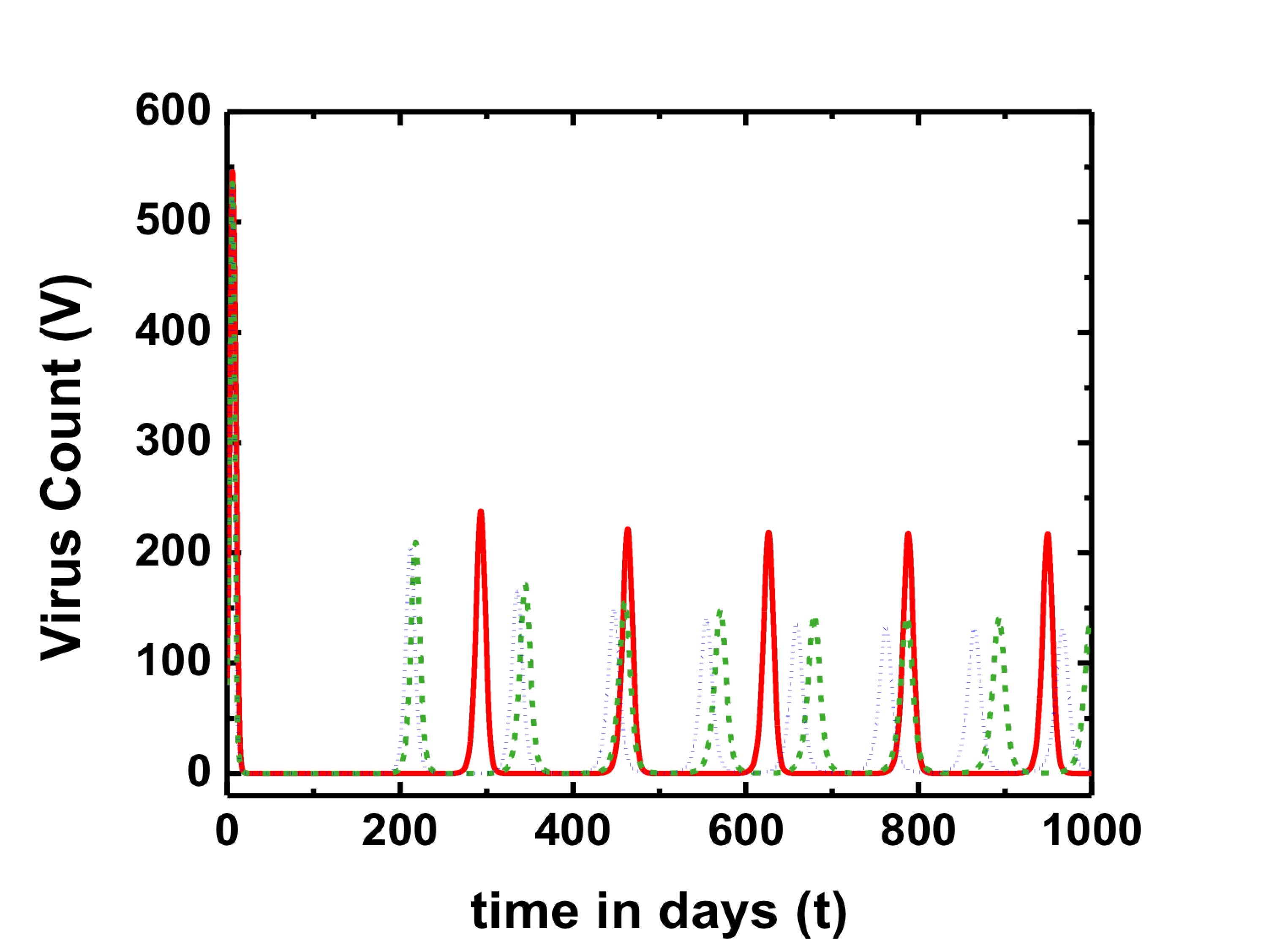}  
\end{center}
\caption{\label{fig:fig7} Time series corresponding to virus particles with  $ \tau_{2}$=0 days (doted curve), $\tau_{2}$=0.3 days (dashed curve) and $\tau_{2}$=4 days (solid curve) for $ f = 0.1 $ and $ k = 5$. The oscillatory behaviour persists for higher $\tau_{2}$ values with increasing amplitude. We choose $ \tau_{1}$=5 days and other parameter values same as for Fig.\ref{fig:fig6}b.}
\end{figure}

The analysis mentioned so far is applicable to the steady states only and does not therefore include the unstable regions shown in Fig.\ref{fig:figa2} and Fig.\ref{fig:fig2b}. Hence for understanding the nature of the dynamics in this region, we numerically analyse the system in equation (\ref{eq:primarymodel}) and find that the dynamic state corresponding to that region is oscillatory. Thus the oscillatory nature of the state which was observed for $ \tau_{2} = 0$ for $f = 0.1 $ and $k = 5$ persists for $\tau_{2} \neq 0$. We also study the change in the dynamics with increase in $\tau_{2}$ as shown in Fig.\ref{fig:fig7}. It is clear that as delay increases, the amplitude of the oscillations increases. This would mean that, large delay in production of antibodies could lead to conditions where the infection recurs periodically as spaced out spikes. 

We know that the actual biological processes considered here occur with some inherent non-zero value of time delay. The analysis given above for three specific cases indicates that the steady states that are stable for $\tau_{2}=0$, can become unstable for a finite time delay $\tau_{2}$ and this method can be applied to all values of parameters to understand the change in stability for each case.

\section{Model Describing Secondary Infection}
\label{sec:4}
The model for secondary infection introduced here is an extension of the primary infection model given in the previous sections. Added to the primary model is a sixth variable which describes the level of  antibody formed in the previous dengue infection that still circulate in the body but are heterologous to the new virus serotype.

In this case the heterologous antibodies can bind to the virions, but the extent of binding not only depends on the complementarity between the antigen and antibody but also on the concentrations of the heterologous antibodies. It is found that different serotypes of virus in the primary and secondary infections have antigenic similarities in their protein coat \cite{Mur11,Hal09}. As a result, the intensity with which a particular virus stimulates a response for heterologous or homologous antibodies will depend on this degree of similarity \cite{Zho06}. Our model introduced two features, correlation factor and dynamic switch,  that can take care of the specific dynamical processes in secondary infection.

We introduce a parameter \textit{w} to specifically take care of the relatedness between the various virus serotypes of primary and secondary infection. This correlation factor \textit{w} can vary from 0 to 1 and it quantifies the similarity between the individual serotypes. 

\begin{figure}[h]
\begin{center}
\includegraphics[width=0.8\textwidth]{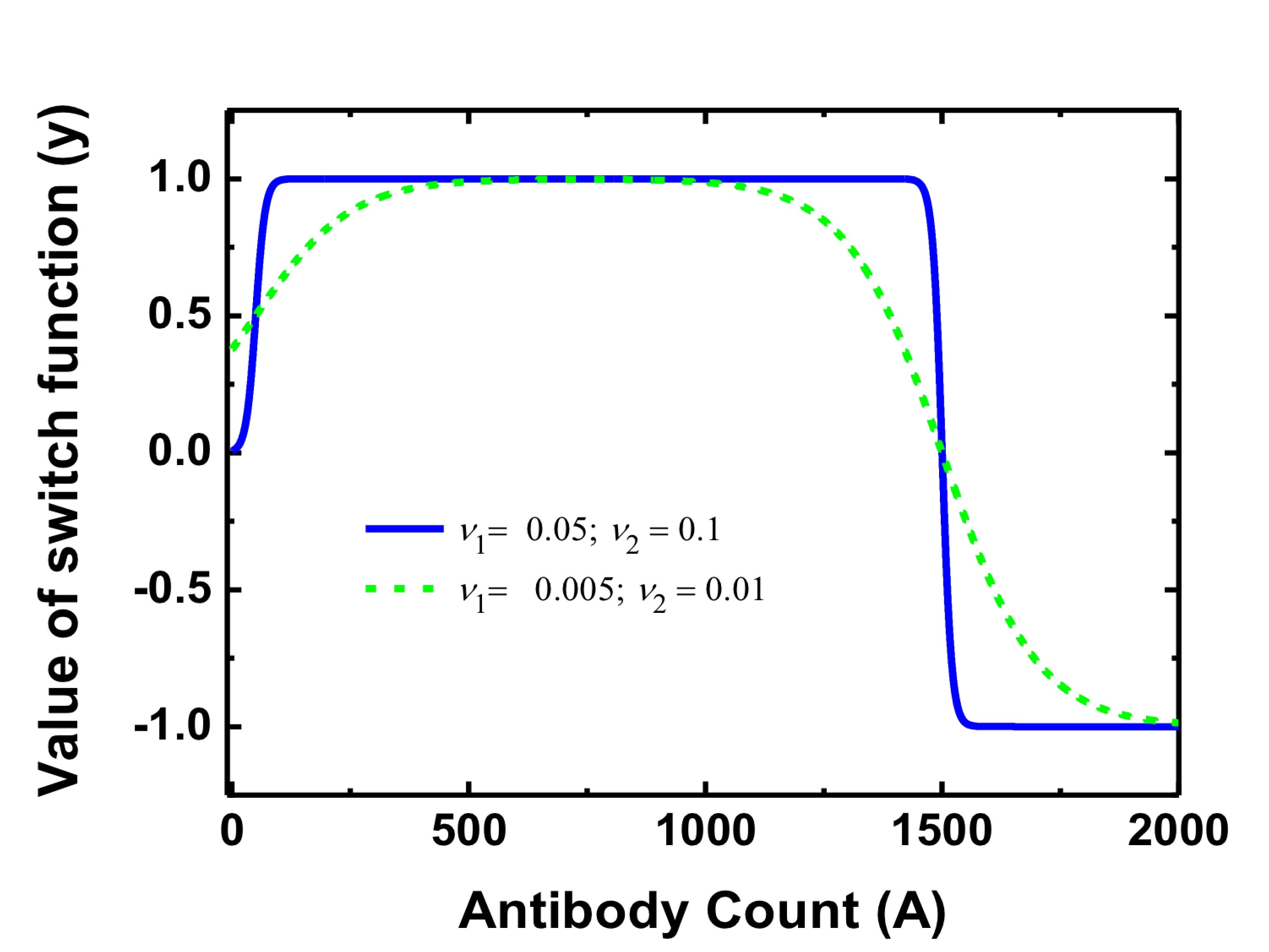}
\end{center}
\caption{\label{fig:figa4} Comparison of $y$ function for $\upsilon_{1}=0.05$, $\upsilon_{2}=0.1$ and $\upsilon_{1}=0.005$, $\upsilon_{2}=0.01$ for $A_{min}=50$ and $A_{max}=1500$. With smaller values of $\upsilon_{1}$ and $\upsilon_{2}$ the switching action of function $y$ extends over larger range of antibodies $A$.}
\end{figure}

For very low concentrations of the heterologous antibodies in the body, the probability of its interaction with virus particles is almost zero. Lack of infection at low heterologous antibody concentrations has been reported \cite{Mur11,Hal03,Chu13}. At high concentrations of the heterologous antibodies there is a good probability of the virus undergoing complete opsonization with these antibodies. In this case it is very probable that the macrophages succeed in destroying the neutralised virions. Hence at very high concentrations the heterologous antibodies actually help in curbing the infection \cite{Wah11,Cha11}. Therefore, the mechanism of ADE is at work only within a certain range in the concentration of heterologous antibodies which corresponds to sub-neutralising concentrations of these antibodies \cite{Hal88,Mur11,Hal09,Hal70}. This means when heterologous antibodies are present in the body between a certain minimum and maximum value ($A_{min}$, $A_{max}$) they actually help in virion production. This is incorporated in our model by introducing a dynamic, smooth switch function \textquoteleft \textit{y}\textquoteright \vspace{2mm}  which is a function of antibody concentration. The new set of non-linear equations describing secondary dengue infection can then be written as:           
\begin{eqnarray}
  \dot{S} & = & \mu - \alpha S - aSV \nonumber \\  
  \dot{I} & = & aSV - \beta I \nonumber \\ 
  \dot{V} & = & kI - \gamma V -yrA_{1}V - pA_{2}V \nonumber \\ 
  \dot{B} & = & \eta - \delta B + cBV \nonumber \\ 
  \dot{A_{1}} & = & wfH(t - \tau_{2})B(t-\tau_{2}) - qA_{1}V - \kappa A_{1} \nonumber \\    
  \dot{A_{2}} & = & fH(t - \tau_{1})B(t - \tau_{2}) - q'A_{2}V - \kappa A_{2} 
\label{eq:secondarymodel}
\end{eqnarray}
with, 
\begin{equation}
y(A_{1})= \dfrac{tanh(\upsilon_{1}(A_{1} - A_{max}))}{1+e^{-\upsilon_{2}(A_{1}-A_{min})}}
\end{equation}
The variable $A_{1}$ stands for the heterologous antibody previously formed on primary infection and $A_{2}$ for the homologous antibody against the new virus serotype of the secondary infection. $A_{2}$ follows dynamics similar to primary infection. The parameters $\upsilon_{1}$ and $\upsilon_{2}$ are chosen such that the switch function $y(A_{1})$  changes its value smoothly from 0 to -1 as $A_{1}$ becomes greater than $A_{min}$ and to +1 as $A_{1}$ rises above $A_{max}$. The nature of $y$ as a function of antibody count for two different sets of parameters $\upsilon_{1}$ and $\upsilon_{2}$ is shown in Fig.(\ref{fig:figa4}). We observe that for smaller values of $\upsilon_{1}$ and $\upsilon_{2}$, the function $y$ shows a smooth transition over larger range of antibody counts $A$. 

The Heaviside step function corresponding to a time lag of $\tau_{1}$ is not associated with the antibody $A_{1}$ as corresponding plasma cells are already present in the form of memory B-cells. On infection with a different serotype virus at t=0, it takes at least $\tau_{2}$ amount of time for the antibodies to get produced for the first time from the B-cells (memory cells) present at t=0. To take care of this we introduce the Heaviside step function $H(t-\tau_{2})$. The new parameters are, \textquoteleft \textit{r}\textquoteright , the rate at which virus particles are neutralised by $A_{1}$ antibodies and \textquoteleft \textit{q $ '$}\textquoteright , the rate at which $A_{2}$ antibody-virus complex is neutralised and degraded. All other parameters in this model are analogous to the primary model.

We first present the analysis of the dynamics of secondary infection by numerical simulations of the above set of equations. Then we study the effect of the parameters of the dynamic switch and the correlation factor on the outcome of the infection.

\section{Numerical Analysis for the Secondary infection model}
\label{sec:5} 
In the detailed numerical analysis of the secondary infection model, we take $q=q'$ and $\textit{r=p}$, with values of $A_{max}=1500$, $A_{min}=50$ and $A_{1}(0)=800$. The values of $\upsilon_{1}$ and $\upsilon_{2}$ have been chosen to be 0.05 and 0.1 respectively. Here the correlation factor \textit{w} is taken as 0.5 for secondary infection with $\tau_{1}$=3 days and $\tau_{2}$=0.3 days. We choose the initial value of $A_{2}$ to be zero. For all other parameters and initial values we take the same values as in the primary model given in Section 3. This system of equations (\ref{eq:secondarymodel}) is numerically solved by using the Runge-Kutta method for delay differential equations \cite{Lak10}. 

\begin{figure}[h]
\begin{center}
\includegraphics[width=0.8\textwidth]{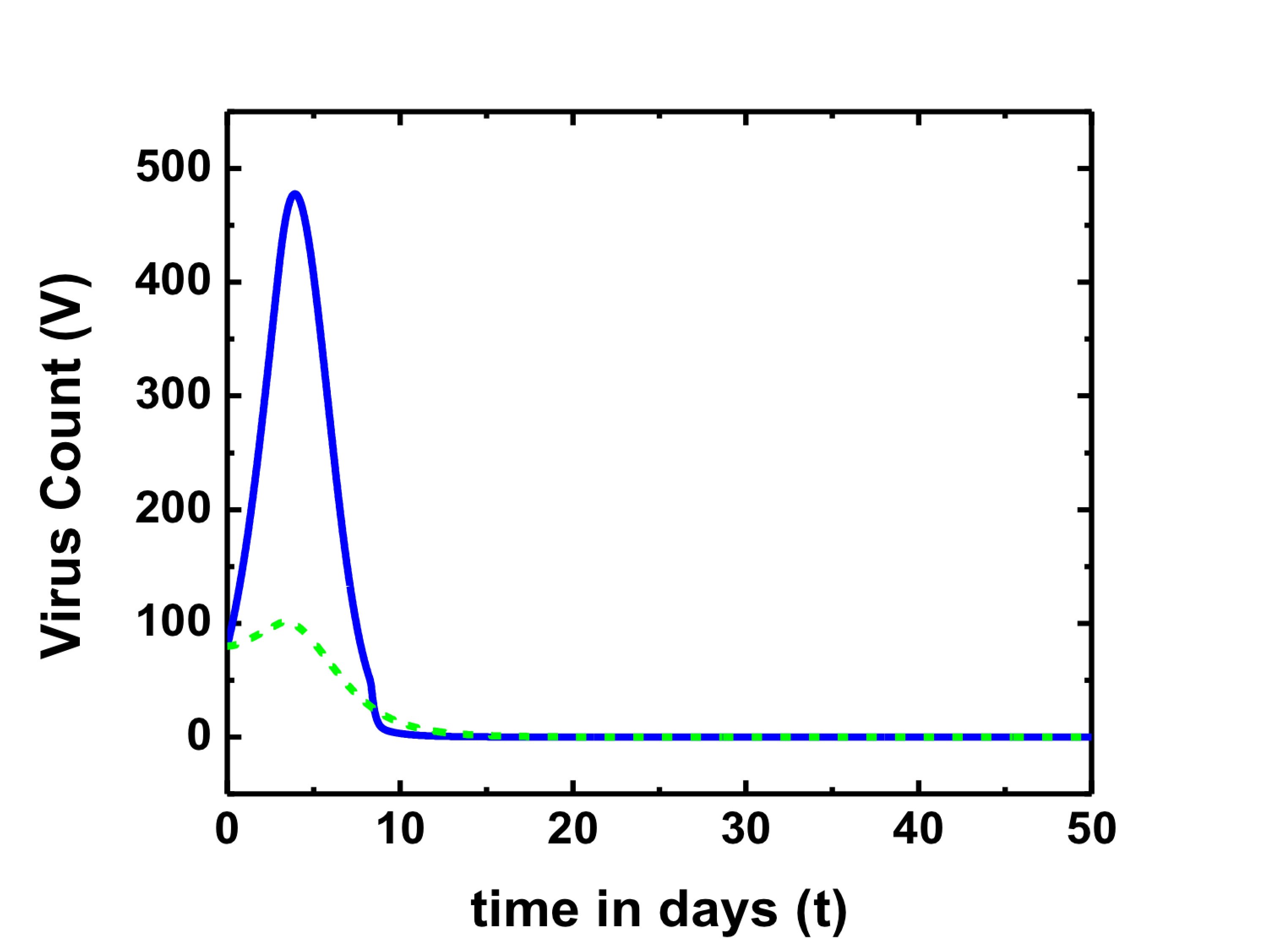}
\end{center}
\caption{\label{fig:fig9} Comparison of virus counts in primary (dashed curve) and secondary (solid curve) infections for $f$=0.8 and $k$=2. We see that the viral count is higher for secondary infection and it gets cleared faster than that in the primary infection. Other parameter values and initial conditions used are given in the text.}
\end{figure}

The viral count in the primary and secondary models obtained from numerical analysis are shown together for comparison in Fig.\ref{fig:fig9}. This indicates that the viral load is cleared from the body within 7 days even for secondary infection and it is faster than the primary infection. This is supported by the reports of observed data, that the viral load is not detected in patients affected by secondary infection when they are suffering from severe conditions like DSS or DHF \cite{Hal88,Hal70}. Moreover, from this figure we observe that the maximum value of viral load is higher in secondary infection as compared to primary infection. Also, clinical data on higher viral loads and faster clearance of viremia in secondary infections has been reported \cite{Vau00}.

\begin{figure}[h]
\begin{center}
\includegraphics[width=0.8\textwidth]{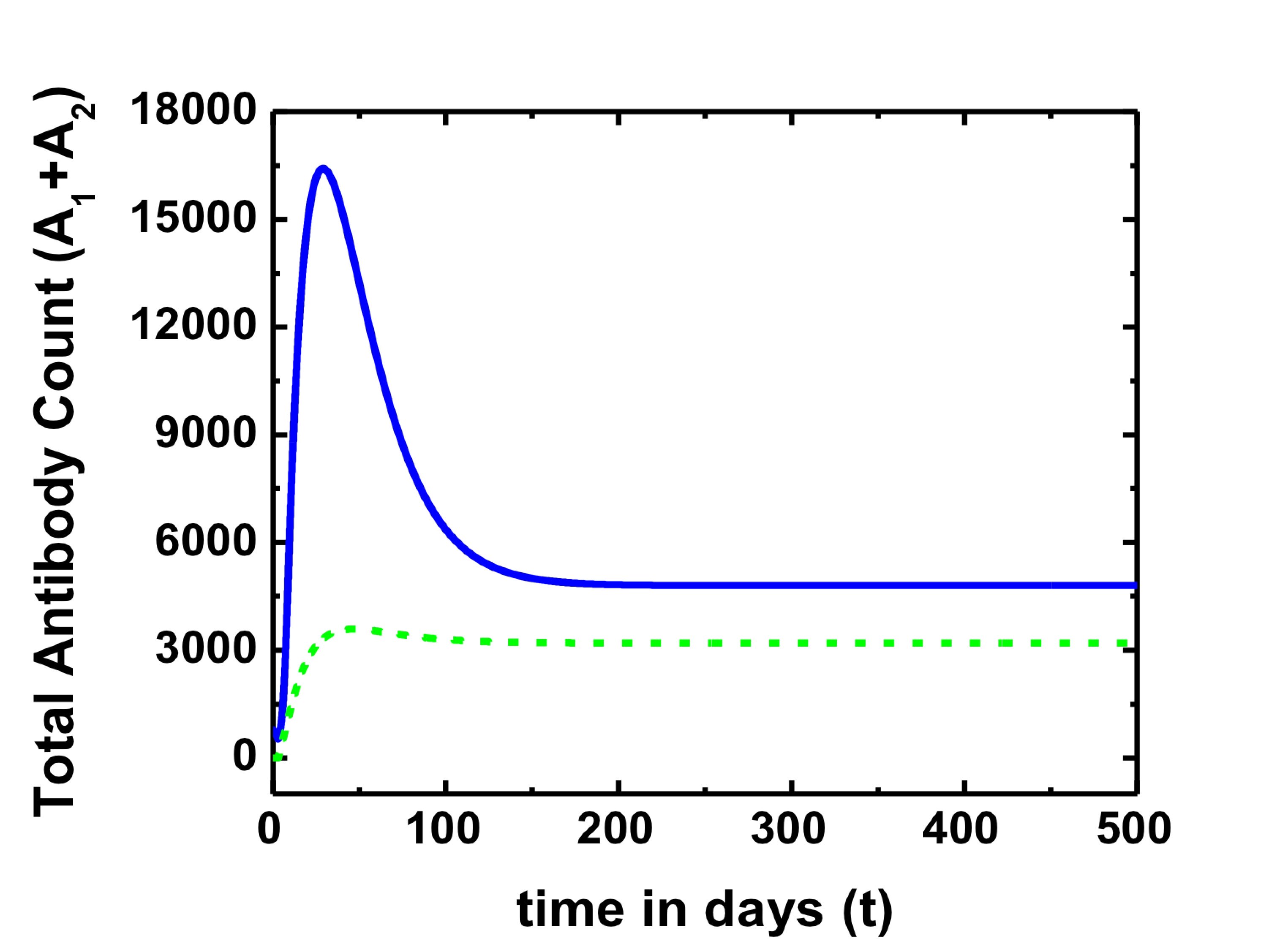}  
\end{center}
\caption{\label{fig:fig10} Total antibody counts during primary (dashed curve) and secondary (solid curve) dengue infections for $f$=0.8 and $k$=2. The total antibody count in secondary infection is more than that in the primary infection.}
\end{figure}

\begin{figure}[h]
\begin{center}
\includegraphics[width=0.8\textwidth]{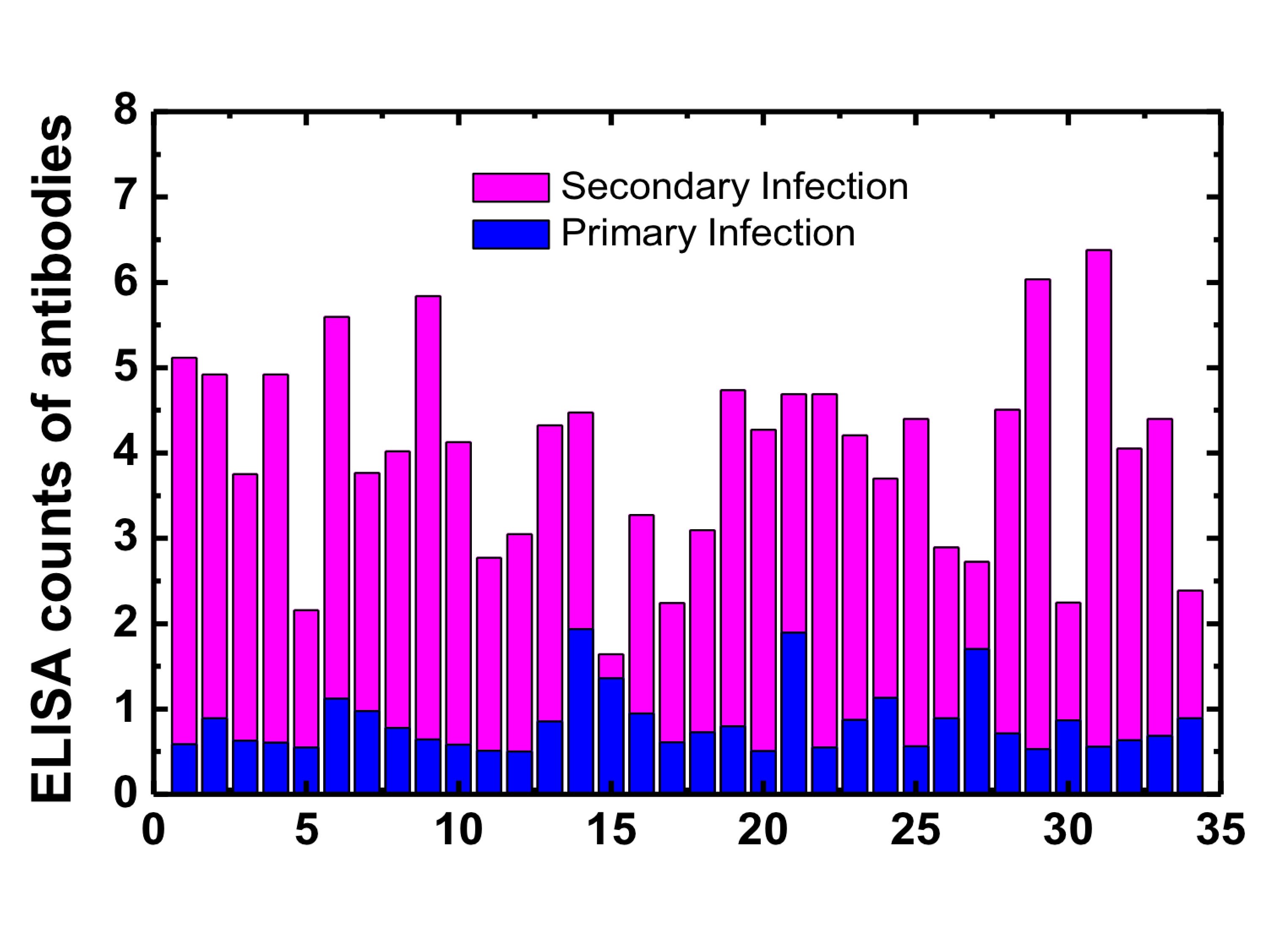} 
\end{center}
\caption{\label{fig:fig11} Clinical data giving total antibodies counts (ELISA count) of 34 patients suffering from primary infection and a separate set of 34 patients suffering from secondary infections. We see that observed antibody counts are higher in the case of secondary infections.}
\end{figure}

One of the important results from the numerical analysis of our secondary infection model is that the total antibody count for secondary infection is much higher than the antibody count in the primary infection. This is shown in Fig.\ref{fig:fig10}. Such stronger immune response caused by higher antibody levels can lead to complex and fatal manifestations resulting in DSS or DHF when infected by a different serotype of DENV as observed in patients with secondary dengue infections \cite{Hal88,Hal70,Fin06,Lei01,Rot99,Rot04}. In support of this, we present the clinical data of the total antibody counts (ELISA counts) taken from a sample set of 34 patients suffering from primary dengue infection and a separate set of 34 patients suffering from secondary infection, supplied by a hospital in Aurangabad, Maharashtra, India in Fig.\ref{fig:fig11}. The data clearly shows that the total antibody counts are higher in secondary dengue infections when compared to primary dengue infection. 


\begin{figure}[h]
\begin{center}
\includegraphics[width=0.8\textwidth]{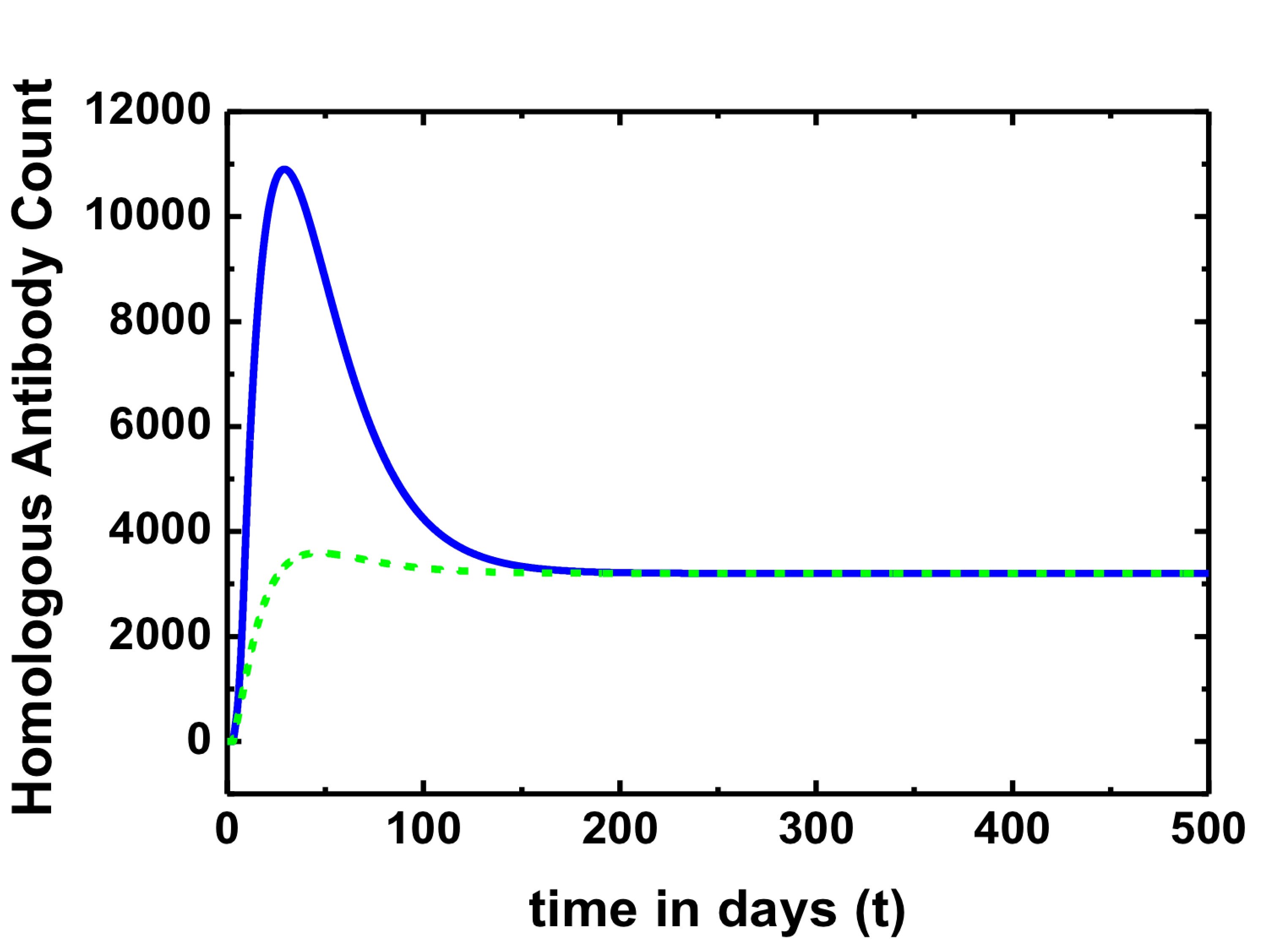}
\end{center}
\caption{\label{fig:fig12} Comparison of homologous antibody count in primary (dashed curve) and secondary (solid curve) infections for $f$=0.8 and $k$=2. The homologous antibody count in the secondary infection is higher compared to that in the primary infection.}
\end{figure}

Another interesting aspect due to the non-linearity of our model is that the production of heterologous antibodies in the secondary infection enhances the production of homologous antibodies. Therefore we observe higher values of homologous antibodies in the secondary infection in comparison to homologous antibodies of the primary infection as is clear from Fig.\ref{fig:fig12}. It leads us to conclude that the higher values of total antibody counts in secondary infection are not only because of the presence of heterologous antibodies but also because of the enhancement in homologous antibody production.

\begin{figure}[h]
\begin{tabular}{cc}
\includegraphics[scale=0.22]{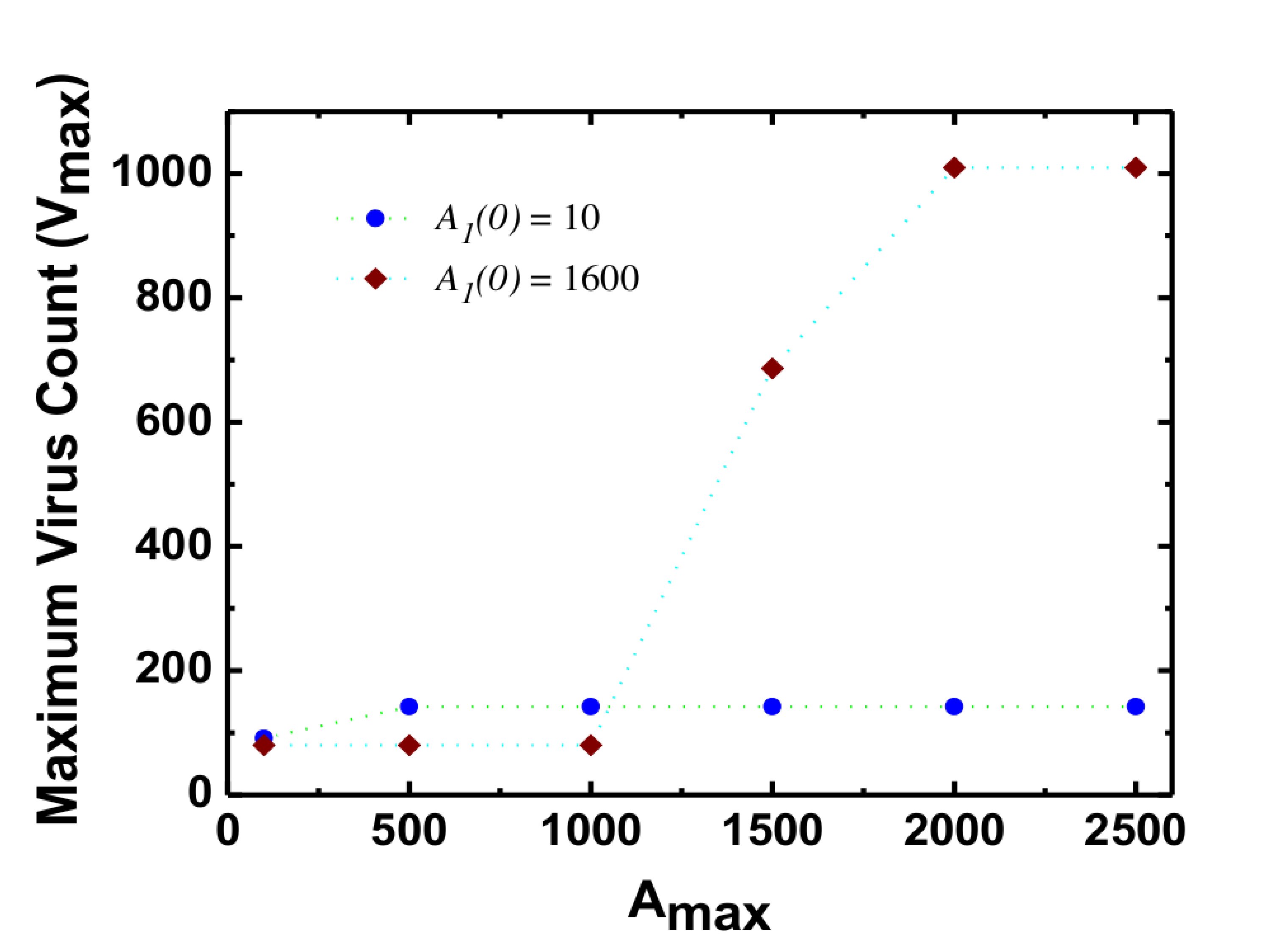}&
\includegraphics[scale=0.22]{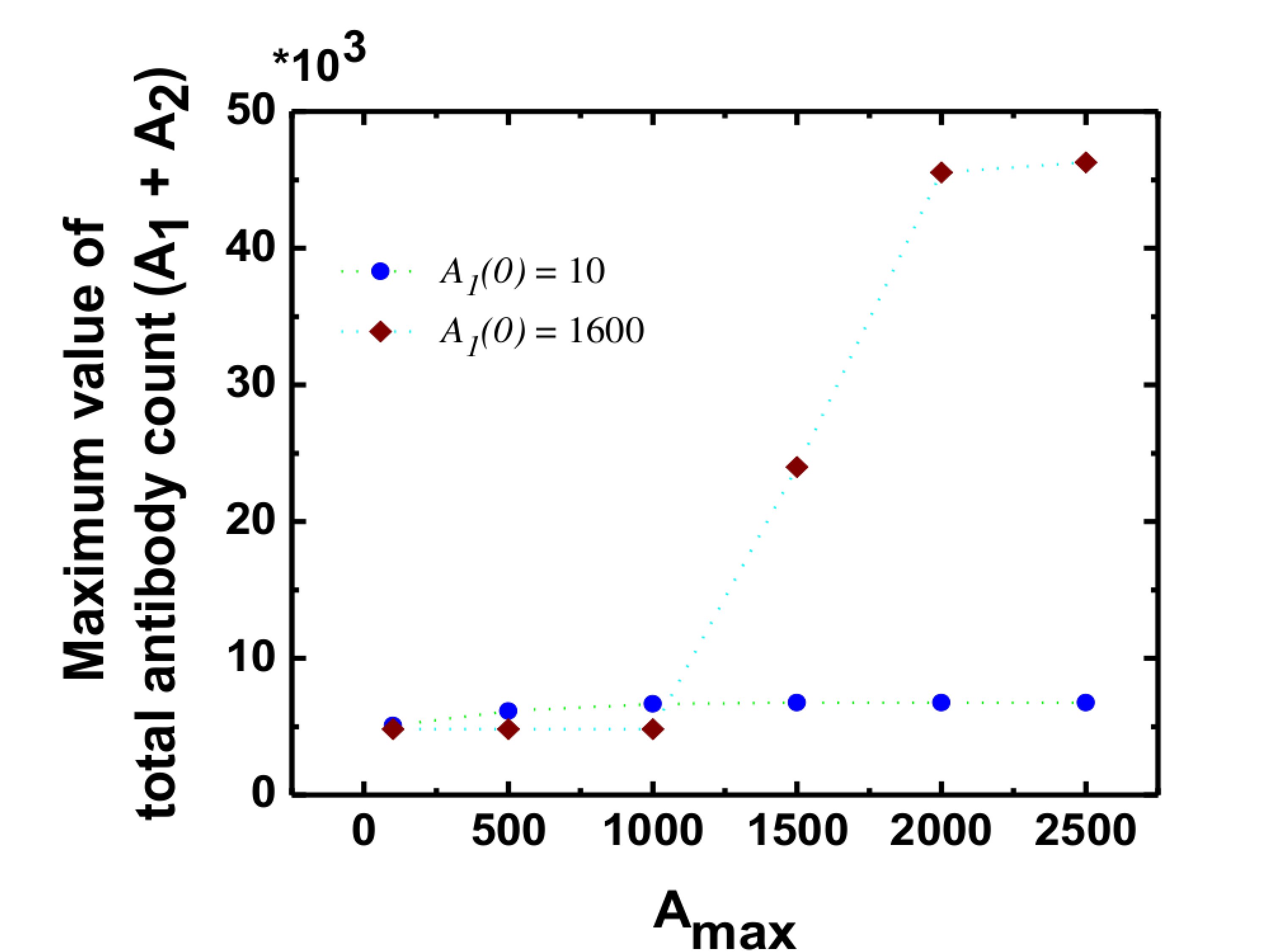}\\
\mbox{\bf {14a}} & \mbox{\bf 14b} \\
\end{tabular}
\begin{center}
\begin{tabular}{c}
\includegraphics[scale=0.22]{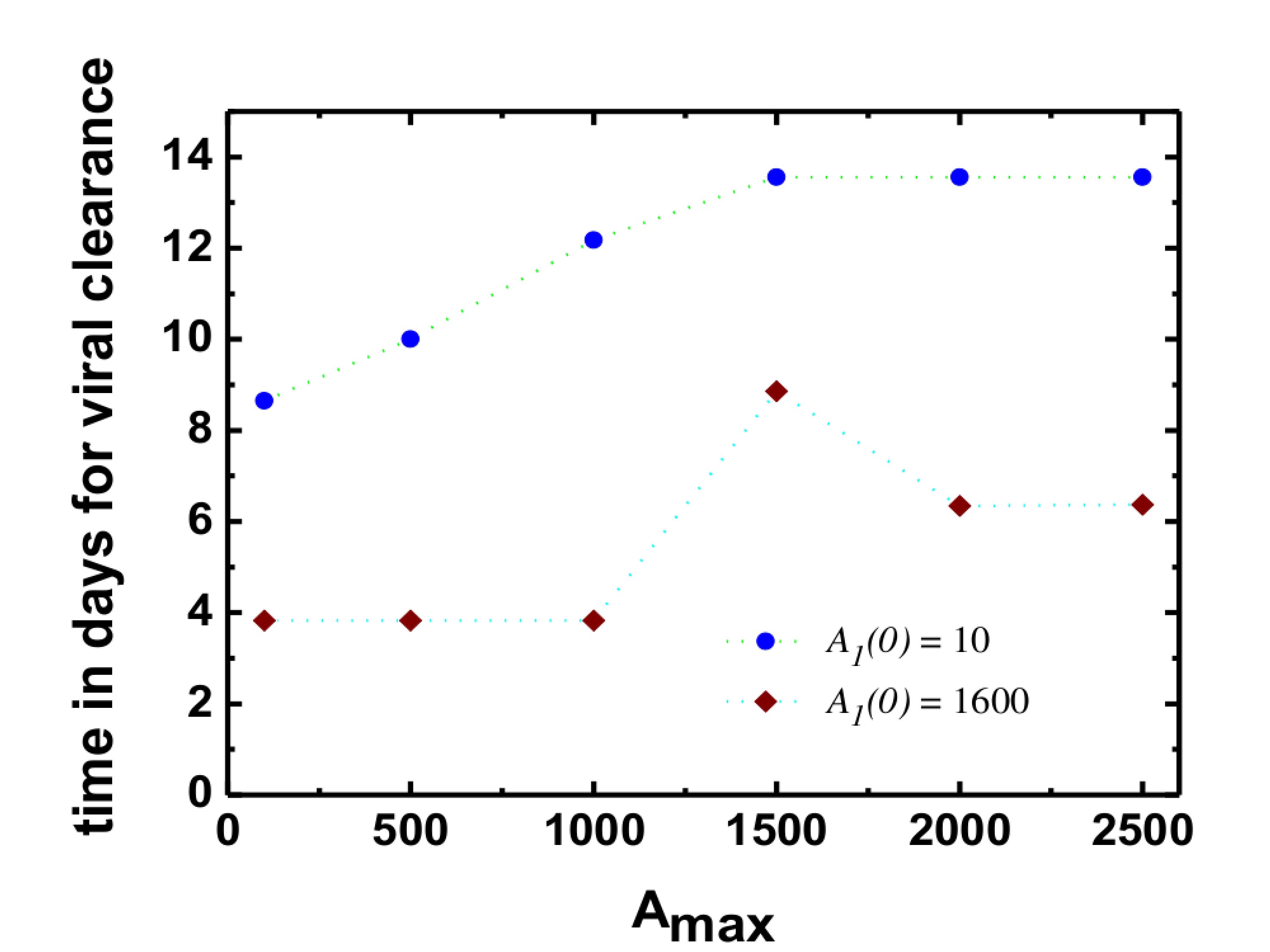}\\
\mbox{\bf 14c}
\end{tabular}
\end{center}

\caption{\label{fig:figa5} Fig.14a The maximum value of virus count calculated for varying $A_{max}$ is shown for two different values of initial heterologous antibody count ($A_{1}(0)$) namely 1600 and 10. Fig14b. The maximum value of total antibody count in the secondary infection is plotted against $A_{max}$ for $A_{1}(0)=$ 10 and 1600. Fig.14c The time taken for viral clearance in secondary infection is plotted against $A_{max}$ for $A_{1}(0)$ =10 and 1600. The parameter value of $f=0.8$ and $k=2$ other parameter values are given in the text.}
\end{figure}

We will now have a look at the effect of various parameters of the switch function $y(A)$ on the outcome of secondary dengue infection in terms of its severity and duration.
For values of $\upsilon_{1}$ and $\upsilon_{2}$ chosen as 0.05 and 0.1 respectively, the maximum value of virus count ($V$) is plotted for different values of $A_{max}$ with $A_{min}=50$. Rest of the parameter values are the same as taken for numerical analysis of the secondary infection mentioned above. It is shown in Fig.(\ref{fig:figa5})a. We observe that when the initial value of heterologous antibody $A_{1}(0)$ is much lower than $A_{min}$, the infection is not very severe and it saturates with increasing $A_{max}$. Such values of maximum virus counts are what we typically get in primary dengue infections. For values of $A_{1}(0)$ larger than $A_{max}$ the infection gets cleared exponentially and the maximum value of the virus count is just the intial $V(0)$ value. For values of $A_{1}(0)$ considerably higher than $A_{min}$ and around $A_{max}$, the maximum value of $V$ increases with increasing $A_{max}$. With further increase in $A_{max}$ the maximum of virus count saturates eventually. Exactly similar observations are made with respect to the maximum of total antibody count ($A_{1}+A_{2}$) plotted for different values of $A_{max}$ as is shown in Fig.(\ref{fig:figa5})b. Thus, we see that the infection aggravates in the range where the ADE mechanism is active as is expected.

Similarly, results for the time taken for the virus count to go below undetectable levels, i.e. the time taken for clearance of infection with varying $A_{max}$ is shown in Fig.(\ref{fig:figa5})c. We have defined the virus clearance threshold to be fixed at $V=10$. For really small values of $A_{1}(0)$, the time taken for the virus to clear is 7-14 days and saturates with increasing $A_{max}$. Again, the values observed are similar to that of what we get in primary dengue infections. For values of $A_{1}(0)$ which are relatively higher than $A_{max}$, the viral clearance is exponential and is cleared in 2-4 days. When $A_{1}(0)$ is around or below $A_{max}$, time taken for viral clearance increases. These values are smaller than the values observed in the case of primary dengue infection as is expected since ADE mechanism is active. 

The above measurements were repeated for $\upsilon_{1}$ and $\upsilon_{2}$ as 0.005 and 0.01 respectively. The results obtained were qualitatively the same. The corresponding values of maximum virus count and maximum total antibody count were found to be slightly smaller (smaller by multiples of ten for virus count and by multiples of few hundreds for total antibody count) in this case. 

\begin{figure}[h]
\begin{tabular}{cc}
\includegraphics[scale=0.22]{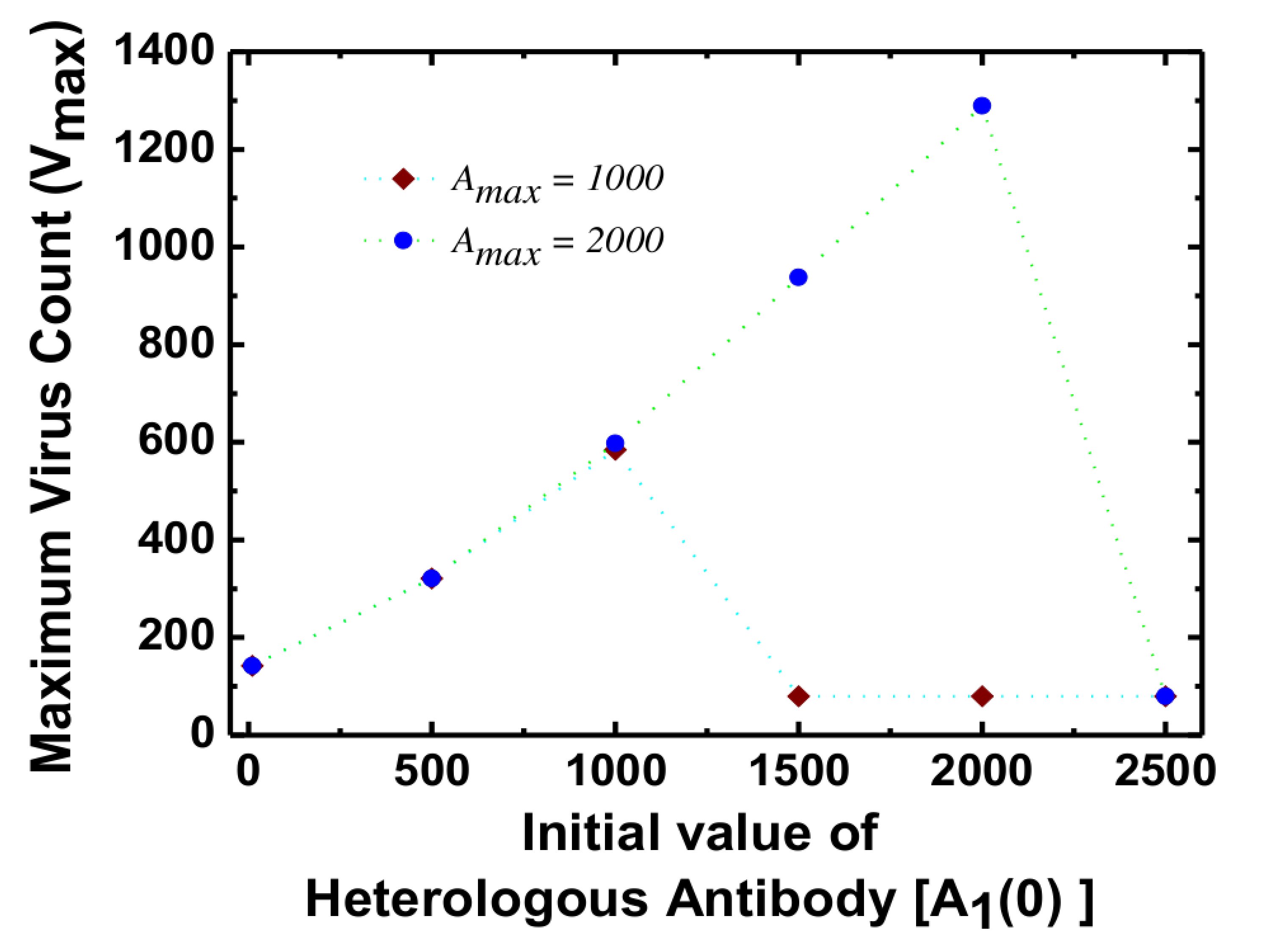}&
\includegraphics[scale=0.22]{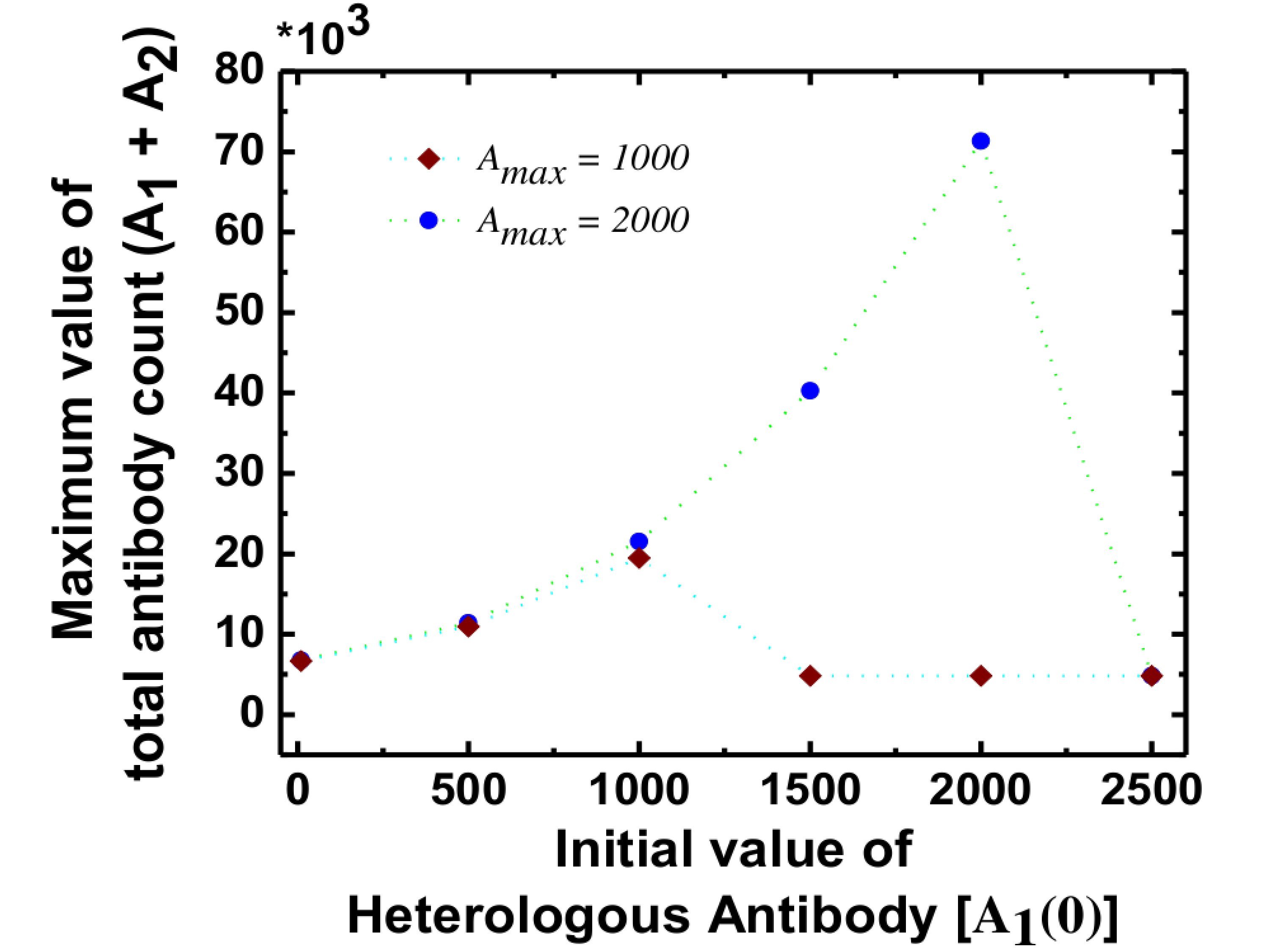}\\
\mbox{\bf {15a}} & \mbox{\bf 15b} \\
\end{tabular}
\begin{center}
\begin{tabular}{c}
\includegraphics[scale=0.22]{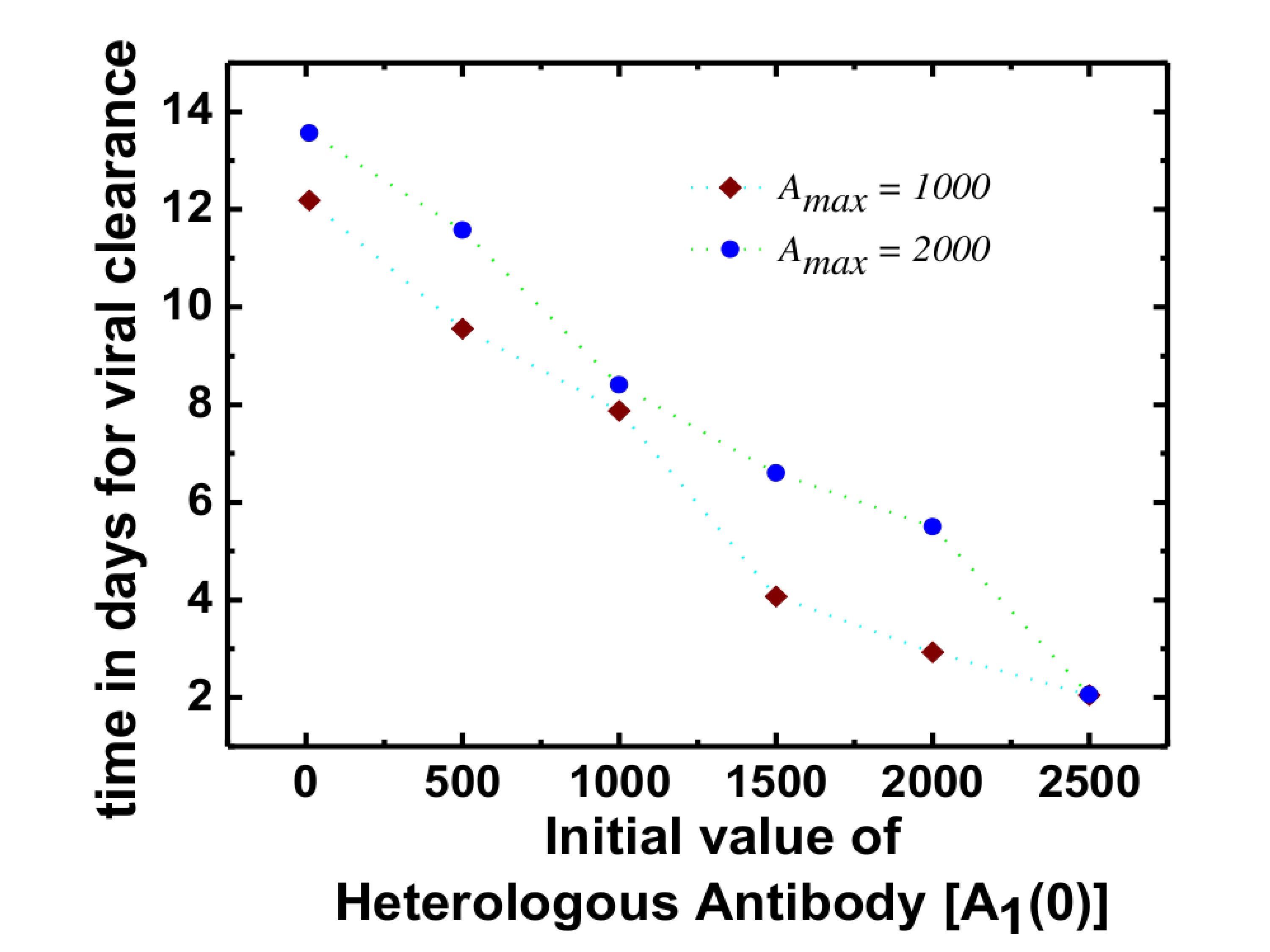}\\
\mbox{\bf 15c}
\end{tabular}
\end{center}
\caption{\label{fig:figa6} Fig.15a The maximum value of virus count calculated for varying initial heterologous antibody count ($A_{1}(0)$) is shown for two different values of $A_{max}$ namely 1000 and 2000. Fig.15b The maximum value of total antibody count in the secondary infection is plotted against $A_{1}(0)$ for $A_{max}=$ 1000 and 2000. Fig.15c The time taken for viral clearance in secondary infection is plotted against $A_{1}(0)$ for $A_{max}$ =1000 and 2000. The parameter value of $f=0.8$ and $k=2$ other parameter values are given in the text.}
\end{figure}

It is known that the concentration of antibodies in the body decays with time \cite{Hal03,Wah11,All07}. Therefore, the initial value of antibodies present from the primary infection would depend on the time interval between the primary and the consecutive secondary dengue infection. Thus, the dependence of the secondary infection severity on this time interval can be studied by looking at the effect of initial heterologous antibody concentration in secondary infection. This study based on the values of maximum viral count, maximum total antibody count and the time taken for viral clearance as discussed above, calculated for different sets of parameter $A_{max}$ is given in Fig.(\ref{fig:figa6}). We have chosen $A_{min}$ as 50 along with $\upsilon_{1}$, $\upsilon_{2}$ as 0.005 and 0.01 respectively. Other parameter values are kept the same as in the above discussion.

From the Fig.(\ref{fig:figa6})a we observe that as the initial concentration of heterologous antibody ($A_{1}(0)$) increases, the maximum viral count increases and peaks when the  value of $A_{1}(0)$ is similar to the value of $A_{max}$. We also note that, higher the value of $A_{max}$ stronger is the peak for maximum viral count. With further increase in $A_{1}(0)$, the virus gets exponentially cleared from the body and the maximum viral count is just the initial viral count. Similar results are observed for the maximum value of the total antibody count as is shown in Fig.(\ref{fig:figa6})b. When time of viral clearance is calculated with increasing $A_{1}(0)$ values we observe that virus gets cleared faster with higher values of $A_{1}(0)$ as is shown in Fig.(\ref{fig:figa6})c. When the $A_{1}(0)$ values are around $A_{max}$ values we observe slight increase in viral clearance time. This is the region where ADE mechanism is most active. For really small values of $A_{1}(0)$ the time taken for viral clearance is as good as what we get in primary infections. Thus, we can conclude that when a host who had suffered from primary dengue infection a long time back, such that the antibodies from the primary infection are negligible in count, gets secondary infection, the infection will not be severe and will be comparable to the primary infection. Similarly, a person who contracts secondary infection just after he has suffered from primary dengue infection, will not face severe consequences as the infection will be exponentially cured. But, if the person were to contract secondary infection after a particular time duration from the primary infection, ADE mechanism would be strongly active and lead to severe infection. This conclusion matches with the observations made in studies based on dengue infection severity with changes in different antibody concentrations \cite{Hal03,Sab59,HalF70,Guz02}.  

\begin{figure}[h]
\begin{center}
\includegraphics[width=0.8\textwidth]{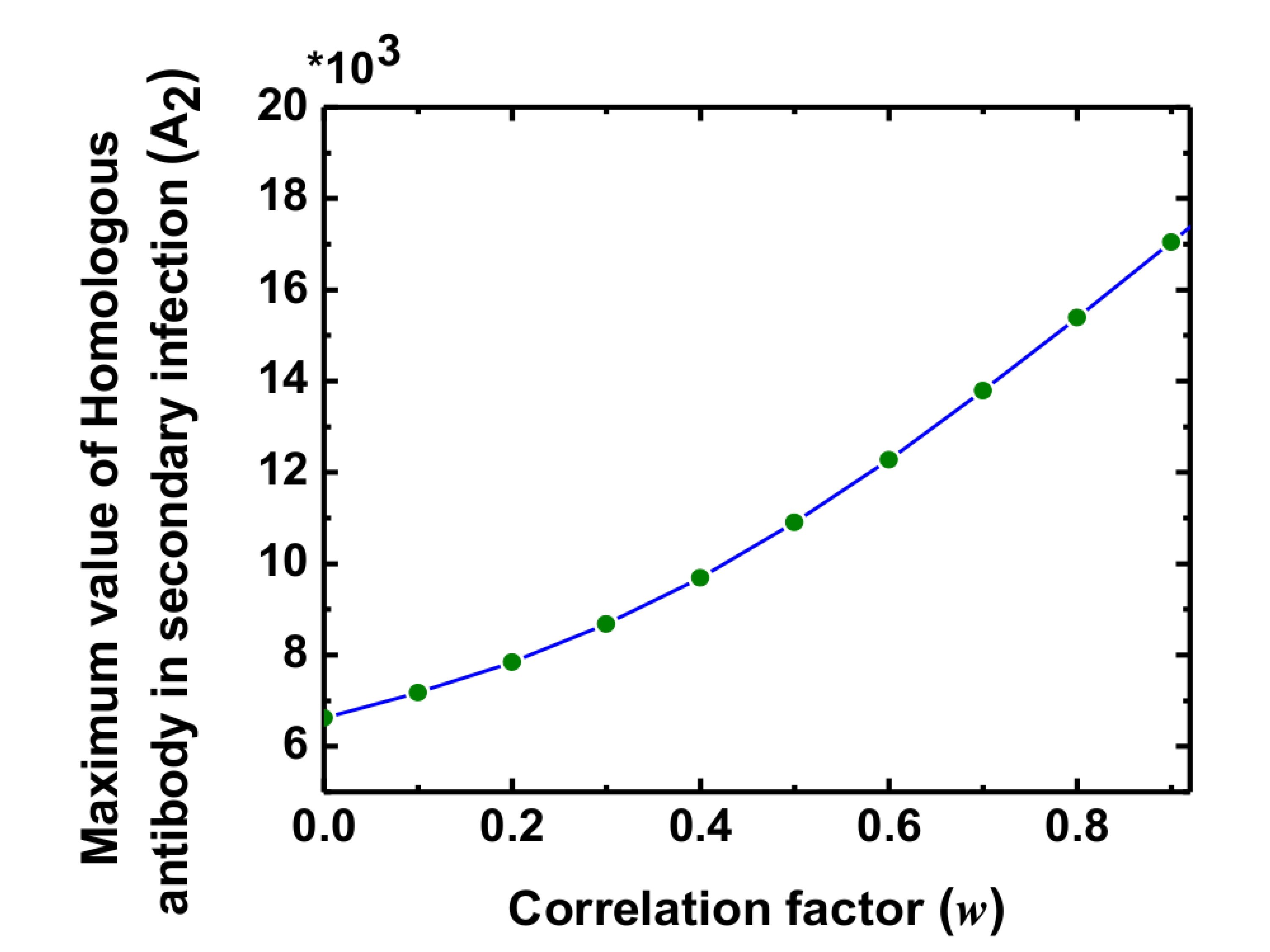} 
\end{center}
\caption{\label{fig:fig13} Maximum homologous antibody count in secondary infection vs. the correlation factor \textit{w} for $f$=0.8 and $k$=2. The maximum homologous antibody count in the secondary infection increases with increasing \textit{w}.}
\end{figure}

We know look at how of the correlation factor $w$ affects the outcome of secondary infection.
In Fig.\ref{fig:fig13} we present the maximum value of homologous antibody count obtained from our numerical analysis as \textit{w} is varied from 0 to 0.9. This shows that the enhancement of homologous antibody production in the secondary infection depends on this factor. There are studies that suggest the severity of secondary dengue infection varies depending on the two DENV serotypes involved \cite{Vau00,Hal70}. In our model we quantify this using the correlation factor \textit{w}. 

\begin{figure}[h]
%\begin{center}
\begin{tabular}{cc}
\includegraphics[scale=1]{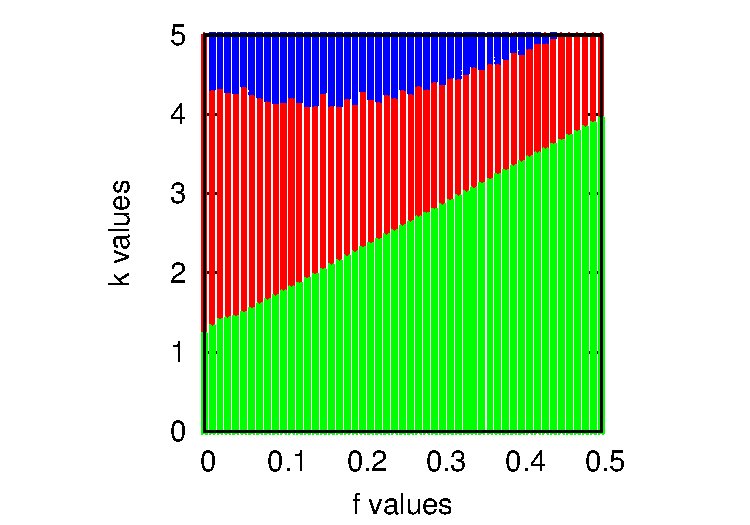} &
\includegraphics[scale=1]{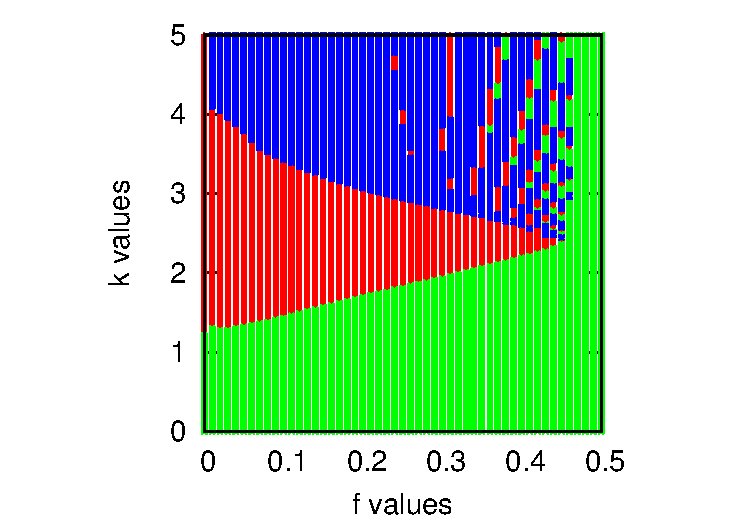} 
\end{tabular}
%\end{center}
\caption{\label{fig:fig8} Stability regions in parameter plane $f$-$k$ for secondary infection. The green(light grey) region corresponds to infection free steady state, the red(dark grey) region corresponds to infectious steady state and the blue(black) region is for oscillatory solutions. The correlation factor $ \textit{w} $ = 0.5(left) and 0.8(right). The other parameter values are given in the text.}
\end{figure}

The regions of stability for possible steady states are identified in the parameter plane $f-k$. This is shown in Fig.\ref{fig:fig8} for two values of \textit{w}, 0.5 and 0.8. It is clear from Fig.\ref{fig:fig8} that in the case for secondary infection also, $V^{*}=0$ state exists which implies that non-infectious steady state is reached. The stability region for this state is shown in green(light grey). We also find the existence of infectious steady state solutions in the region shown in red(dark grey) and oscillatory solutions in region shown in blue(black). Moreover, we note that, as the value of \textit{w} increases the region of stability with zero virus count shifts to lower values of $f$. This implies that even though the severity of the infection might increase with \textit{w} (because of enhanced antibody production) the infection free stability region in fact increases. Further biological studies are needed to understand the nature of this correlation or the degree of similarity between the different serotypes and the severity of infection.   

\section{Discussion}
\label{sec:6}
In the current global context, dengue forms a major concern due to its severity and complexity. Considered to be a tropical disease, it is spreading to higher latitudes due to changes in climatic and human activities. Hence, it is of great importance to study the epidemiology of dengue infections on a global macro-epidemic scale. Here we emphasise the relevance of studying it on a micro-epidemic level to know the mechanism involved in the infection which will possibly help to eradicate it.

In this paper, we present a model based on several observed features of dengue infection. We consider the humoral or antibody mediated immune response as the main mechanism that is involved in viral clearance and subsequent development of immunity since this is reported as the relevant one in the context of dengue \cite{Mur11,Jen08}. It is known that host immune mechanism involves multiple cellular processes which are taken care of by introducing a delay time $\tau_{1}$ in the initial production of antibodies after maturation of naive B-cells to virus specific plasma cells (modified B-cells) and another delay $\tau_{2}$ for the production of antibodies from plasma cells once they are formed. The analysis presented here brings out the dependence of virus counts on delays and the specific range of parameters for which infection free state is stable.

Our results imply that in general the virus gets cleared in 7-14 days which is in agreement with clinical literature. An interesting outcome of the primary infection model is a critical value $f_{c}$ for the parameter $f$ representing immune response, such that when $f>f_{c}$ we always get infection free stable state which corresponds to what is observed in this context. However, for values of $f<f_{c}$, which means that immune response is weak and burst rate of virus is large, it is possible that the virus count will settle to a non-zero value, corresponding to infectious steady state. Our analysis also brings out the dependence of stability of solutions for the primary model on the delay parameter $\tau_{2}$. Specifically, if the delay in the immune response is considerable, infectious steady state can become recurring peaks of infection. 

The dynamics of secondary infection with a different DENV serotype is modelled to take care of the ADE process and the consequent production of higher viremia and antibodies which can lead to severe manifestations of DHF and DSS are the obtained outcomes of the model. Since the mechanism is found to be dependent on the concentrations of heterologous antibodies present from previous primary infection, we have introduced a dynamic switch function which can take care of this dependence. The effects of varying the parameters of the switch function on the severity and duration of the secondary infection are studied in detail. Our results also support the observed clinical data on increased antibody counts in secondary infection. The degree of similarity between the two virus serotypes involved is a crucial factor that decides the ability of the heterologous antibodies to neutralise viral particles. In our model the correlation factor \textit{w} is introduced as a measure of this similarity. Our results suggests that the antibody production and correspondingly the severity of secondary infection depends on this similarity.          

In conclusion we note that this is the first attempt to model primary dengue infection along with the secondary infection on a micro-epidemic level using humoral immune response. The delay in the immune response, the correlation factor and the dynamic switch in the secondary infection model are the important features of this work. The results obtained so forth are very encouraging to further extend this work for the mathematical analysis of the secondary infection model and tune it to the fine details of the unusual virus-host interactions observed in dengue infections with biological data. Further work in this direction will surely help in understanding the complex mechanisms involved in dengue pathogenesis. It will also be helpful to formulate relevant schemes to develop a dengue vaccine which can provide immunity against all four serotypes in future.

\begin{acknowledgements}
We would like to thank Dr. Hedgewar Hospital, Aurangabad, Maharashtra for providing the ELISA data on Dengue infections. We thank the anonymous reviewers for their valuable comments.
\end{acknowledgements}
\section*{Appendix\\
Details of the Jacobian and the Characteristic Equation} 
The Jacobian describing the primary infection in terms of $S^{*}$, $V^{*}$, $B^{*}$ and $A^{*}$ which are the equilibrium values of healthy cells, virus count, B-cells and antibody count respectively is given as,
\begin{equation}
 J = \left[ \begin{array}{ccccc}
-(\alpha+aV^{*}) & 0 & -aS^{*} & 0 & 0\\
aV^{*} & -\beta & aS^{*} & 0 & 0\\
0 & k & -(\gamma+pA^{*}) & 0 & -pV^{*}\\                                             
0 & 0 & cB^{*} & -(\delta-cV^{*}) & 0\\
0 & 0 & -qA^{*} & fe^{-\lambda\tau_{2}} & -(\kappa+qV^{*})\\
\end{array} \right] 
\label{eq:eq1}
\end{equation} 
where;
\begin{subequations}
\label{eq:eq2}
\begin{eqnarray}
  S^{*} & = & \frac{\mu}{(\alpha + aV^{*})}  \\  
  I^{*} & = & \frac{aS^{*} V^{*}}{\beta}  \\
  B^{*} & = & \frac{\eta}{(\delta - cV^{*})}\\
  A^{*} & = & \frac{f B^{*}}{( qV^{*} + \kappa)}  
\end{eqnarray}
\end{subequations}
The characteristic equation for the Jacobian $J$ with eigenvalues $\lambda$ can be written as:
\begin{equation}
\label{eq:eqn3}
G(\lambda)=\lambda^{5} + G_{4}\lambda^{4} + G_{3}\lambda^{3} + G_{2}\lambda^{2} + G_{1}\lambda^{1} + G_{0} \lambda + H_{2}e^{-\lambda\tau_{2}}\lambda^{2} + H_{1}e^{-\lambda\tau_{2}}\lambda + H_{0}e^{-\lambda\tau_{2}} = 0
\end{equation}                                                        
Where:
\begin{IEEEeqnarray}{rCl}
G_{4} & = & ( \alpha + \beta + \gamma + \delta + \kappa + V^{*}(a + \kappa -c) + pA^{*}); \nonumber \\
G_{3} & = & ( \alpha\beta + \alpha\gamma + \gamma\beta + \alpha\delta + \beta\delta +\gamma\delta + \alpha\kappa + \beta\kappa + \gamma\kappa + \delta\kappa \nonumber \\
&& + \: V^{*}(a( \beta + \gamma + \delta + \kappa ) + q( \alpha+ \beta + \gamma + \delta ) - c( \alpha + \beta + \gamma + \kappa )) + V^{*2}(qa-ac-cq) \nonumber \\
&& + \: A^{*}V^{*}p(a-c) + A^{*}p(\alpha + \beta + \delta + \kappa) - akS^{*}); \nonumber\\
G_{2} & = & ( \alpha\beta\gamma + \alpha\beta\delta + \alpha\gamma\delta + \beta\gamma\delta + \alpha\beta\kappa + \alpha\gamma\kappa + \beta\gamma\kappa + \alpha\kappa\delta + \beta\delta\kappa + \gamma\delta\kappa  \nonumber \\
&& + \: V^{*}(a( \beta\gamma + \beta\delta + \gamma\delta + \beta\kappa + \gamma\kappa + \delta\kappa ) + q(\alpha\beta + \alpha\gamma + \beta\gamma + \alpha\delta + \beta\delta + \gamma\delta) \nonumber \\
&& - \: c(\alpha\beta + \alpha\gamma + \beta\gamma + \alpha\kappa + \beta\kappa + \gamma\kappa)) \nonumber \\
&& + \: V^{*2}(aq\beta + aq\gamma + aq\delta - c(a\beta + a\gamma + q\alpha + a\kappa + q\beta + q\gamma)) - aqcV^{*3} \nonumber \\
&& + \: A^{*}p(\alpha\beta + \alpha\delta + \beta\delta + \gamma\kappa + \beta\kappa + \delta\kappa) + A^{*}V^{*}p(a(\beta + \delta + \kappa) - c(\alpha + \beta + \kappa)) \nonumber \\ 
&& - \: apcA^{*}V^{*2} - akS^{*}(\kappa + \alpha + \delta + qV^{*} -cV^{*}));  \nonumber\\
G_{1} & = & ( \alpha\beta\gamma\delta + \alpha\beta\gamma\kappa + \alpha\beta\delta\kappa + \alpha\gamma\delta\kappa + \beta\gamma\delta\kappa \nonumber \\
&& +\: V^{*}(a(\beta\gamma\delta + \beta\gamma\kappa + \beta\delta\kappa + \gamma\delta\kappa) + q(\alpha\beta\gamma + \alpha\beta\delta + \alpha\gamma\delta + \beta\gamma\delta) \nonumber \\
&& - \: c(\alpha\beta\gamma + \alpha\beta\kappa + \alpha\gamma\kappa + \beta\gamma\kappa)) \nonumber\\
&& +\: V^{*2}(aq(\beta\gamma + \beta\delta + \gamma\delta) - cq(\alpha\beta + \alpha\gamma + \beta\gamma) - ac(\beta\gamma + \beta\kappa + \gamma\kappa)) -aqcV^{*3}(\beta + \gamma) \nonumber\\
&& +\: A^{*}p(\alpha\beta\delta + \alpha\beta\kappa + \alpha\delta\kappa + \beta\delta\kappa) + A^{*}V^{*}p(a(\beta\delta + \beta\kappa + \delta\kappa) - c(\alpha\beta + \alpha\kappa + \beta\kappa)) \nonumber \\ 
&& -\: A^{*}V^{*2}apc(\beta + \kappa) - akS^{*}(\alpha\kappa + \delta\kappa + \alpha\delta + V^{*}(q\alpha + q\delta -c\kappa -c\alpha -cqV^{*})));  \nonumber\\
G_{0} & = & (\alpha\beta\gamma\delta\kappa + V^{*}(q\alpha\beta\gamma\delta + k\alpha\beta\gamma\delta - c\alpha\beta\gamma\kappa) \nonumber\\
&& +\: V^{*2}(aq\beta\gamma\delta -cq\alpha\beta\gamma -ac\beta\gamma\kappa) - V^{*3}acq\beta\gamma + A^{*}p\alpha\beta\kappa\delta  \nonumber \\
&& +\: A^{*}V^{*}p(q\alpha\beta\delta + a\beta\delta\kappa - c\alpha\beta\kappa - q\alpha\beta\delta) - A^{*}V^{*2}apc\beta\kappa  \nonumber \\
&& -\: akS^{*}(\alpha\delta\kappa + a\alpha\delta V^{*} - c\alpha\kappa V^{*} - cq\alpha V^{*2})); \nonumber\\
H_{2} & = & (cfpB^{*}V^{*}); \nonumber\\
H_{1} & = & cfp(\alpha B^{*}V^{*} + \beta B^{*}V^{*} + aB^{*}V^{*2}); \nonumber\\
H_{0} & = & cfp\beta (\alpha B^{*}V^{*} + aB^{*}V^{*2}) 
\end{IEEEeqnarray}

For $V^{*}$ = $0$, we get the coefficients of the characteristic equation (\ref{eq:eqn3}) as:

\begin{IEEEeqnarray}{rCl}
G_{4} & = & ( \alpha + \beta + \gamma + \delta + \kappa + pA^{*}); \nonumber \\
G_{3} & = & ( \alpha\beta + \alpha\gamma + \gamma\beta + \alpha\delta + \beta\delta +\gamma\delta + \alpha\kappa + \beta\kappa + \gamma\kappa + \delta\kappa \nonumber \\
&& + \:  A^{*}p(\alpha + \beta + \delta + \kappa) - akS^{*}); \nonumber\\
G_{2} & = & ( \alpha\beta\gamma + \alpha\beta\delta + \alpha\gamma\delta + \beta\gamma\delta + \alpha\beta\kappa + \alpha\gamma\kappa + \beta\gamma\kappa + \alpha\kappa\delta + \beta\delta\kappa + \gamma\delta\kappa  \nonumber \\
&& + \: A^{*}p(\alpha\beta + \alpha\delta + \beta\delta + \gamma\kappa + \beta\kappa + \delta\kappa) - akS^{*}(\kappa + \alpha + \delta ));  \nonumber\\
G_{1} & = & ( \alpha\beta\gamma\delta + \alpha\beta\gamma\kappa + \alpha\beta\delta\kappa + \alpha\gamma\delta\kappa + \beta\gamma\delta\kappa \nonumber \\
&& +\: A^{*}p(\alpha\beta\delta + \alpha\beta\kappa + \alpha\delta\kappa + \beta\delta\kappa) - akS^{*}(\alpha\kappa + \delta\kappa + \alpha\delta ));  \nonumber\\
G_{0} & = & (\alpha\beta\gamma\delta\kappa +  A^{*}p\alpha\beta\kappa\delta - akS^{*}(\alpha\delta\kappa)); \nonumber\\
H_{2} & = & 0; \nonumber\\
H_{1} & = & 0; \nonumber\\
H_{0} & = & 0 
\end{IEEEeqnarray}

On substituting the above coefficients in equation (\ref{eq:eqn3}) and simplifying after using equation (\ref{eq:eq2}) we get the characteristic equation for $V^{*}$ = $0$ as

\begin{equation}
G(\lambda) = (\alpha + \lambda)(\kappa + \lambda)(\delta + \lambda)(\lambda^{2} + \lambda(\gamma(1 + \frac{fp\eta}{\delta \gamma \kappa}) + \beta) + \beta \gamma(1 + \frac{fp\eta}{\delta \gamma \kappa} - R_{0}))
\label{eq:chareqvnew}
\end{equation}

where;
\begin{equation}
R_{0}=\frac{a\mu k}{\alpha\beta\gamma} \nonumber
\end{equation}
We carry out further analysis in the paper by considering the coefficient values given in this Appendix.

\end{document}